\documentclass[preprint,journal]{vgtc}            

\onlineid{1240}

\ieeedoi{10.1109/TVCG.2023.3326940}

\vgtccategory{Research}

\vgtcpapertype{application}

\newcommand{\paperTitle}{SkiVis: Visual Exploration and Route Planning in Ski Resorts}
\title{\paperTitle}

\author{%
    \authororcid{Julius Rauscher}{0000-0003-1318-9642},
    \authororcid{Raphael Buchmüller}{0000-0002-0612-8828},
    \authororcid{Daniel A. Keim}{0000-0001-7966-9740},
    and \authororcid{Matthias Miller}{0000-0002-6281-2173}
}

\authorfooter{
  \item
  	Julius Rauscher, Raphael Buchmüller, Daniel A. Keim and Matthias Miller are with University of Konstanz.\\
  	E-mail: firstname.lastname@uni-konstanz.com
}

\abstract{%
Optimal ski route selection is a challenge based on a multitude of factors, such as the steepness, compass direction, or crowdedness.
The personal preferences of every skier towards these factors require individual adaptations, which aggravate this task.
Current approaches within this domain do not combine automated routing capabilities with user preferences, missing out on the possibility of integrating domain knowledge in the analysis process.
We introduce \skivis, a visual analytics application to interactively explore ski slopes and provide routing recommendations based on user preferences. 
In collaboration with ski guides and enthusiasts, we elicited requirements and guidelines for such an application and propose different workflows depending on the skiers' familiarity with the resort.
In a case study on the resort of Ski Arlberg, we illustrate how to leverage volunteered geographic information to enable a numerical comparison between slopes.
We evaluated our approach through a pair-analytics study and demonstrate how it supports skiers in discovering relevant and preference-based ski routes.
Besides the tasks investigated in the study, we derive additional use cases from the interviews that showcase the further potential of \skivis,
\jr{and contribute directions for further research opportunities.} 

}

\keywords{Geographic Visualization, Routing.}

\usepackage[english]{babel}

\usepackage[svgnames]{xcolor}
\usepackage{pdfrender}
\usepackage[skins]{tcolorbox}
\usepackage{tikz}
\usetikzlibrary{shapes.geometric}
\usepackage{microtype}
\usepackage{wrapfig}
\usepackage{makecell}
\usepackage{enumitem}
\usepackage{caption}
\usepackage{subcaption}
\usepackage[pagebackref,bookmarks]{hyperref}
\addto\extrasenglish{
  
}
\hypersetup{
    unicode=false,          
    pdftoolbar=true,        
    pdfmenubar=true,        
    pdffitwindow=false,     
    pdfstartview={FitH},    
    pdftitle={My title},    
    pdfauthor={Author},     
    pdfsubject={Subject},   
    pdfcreator={Creator},   
    pdfproducer={Producer}, 
    pdfkeywords={keyword1, key2, key3}, 
    pdfnewwindow=true,      
    colorlinks=true,       
    filecolor=magenta,      
    urlcolor=black           
}
  
\newcommand{\subhead}[1]{\vspace{2pt} \noindent \textbf{#1}}
\newcommand{\subheadnvs}[1]{\noindent \textbf{#1}}

\definecolor{c_orange}{HTML}{ffa500}

\newcommand{\bluecircle}{\tikz\draw[blue,fill=blue] (0,0) circle (.5ex);}
\newcommand{\redcircle}{\tikz\draw[red,fill=red] (0,0) circle (.5ex);}
\newcommand{\blackcircle}{\tikz\draw[black,fill=black] (0,0) circle (.5ex);}
\newcommand{\orangecircle}{\tikz\draw[c_orange,fill=c_orange] (0,0) circle (.5ex);}

\newcommand{\reddiamond}{\begin{tikzpicture} \node[diamond,draw=red,fill=red, scale=0.45] (d) at (0,0) {}; \end{tikzpicture}}
\newcommand{\blackdiamond}{\begin{tikzpicture} \node[diamond,draw=black,fill=red, scale=0.45] (d) at (0,0) {}; \end{tikzpicture}}

\newcommand{\heartwhite}{%
  \begingroup\normalfont
  \includegraphics[height=1.3\fontcharht\font`\B]{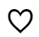}%
  \endgroup
}
\newcommand{\heartblack}{%
  \begingroup\normalfont
  \includegraphics[height=1.3\fontcharht\font`\B]{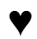}%
  \endgroup
}
\newcommand{\tBarLift}{%
  \begingroup\normalfont
  \includegraphics[height=1.3\fontcharht\font`\B]{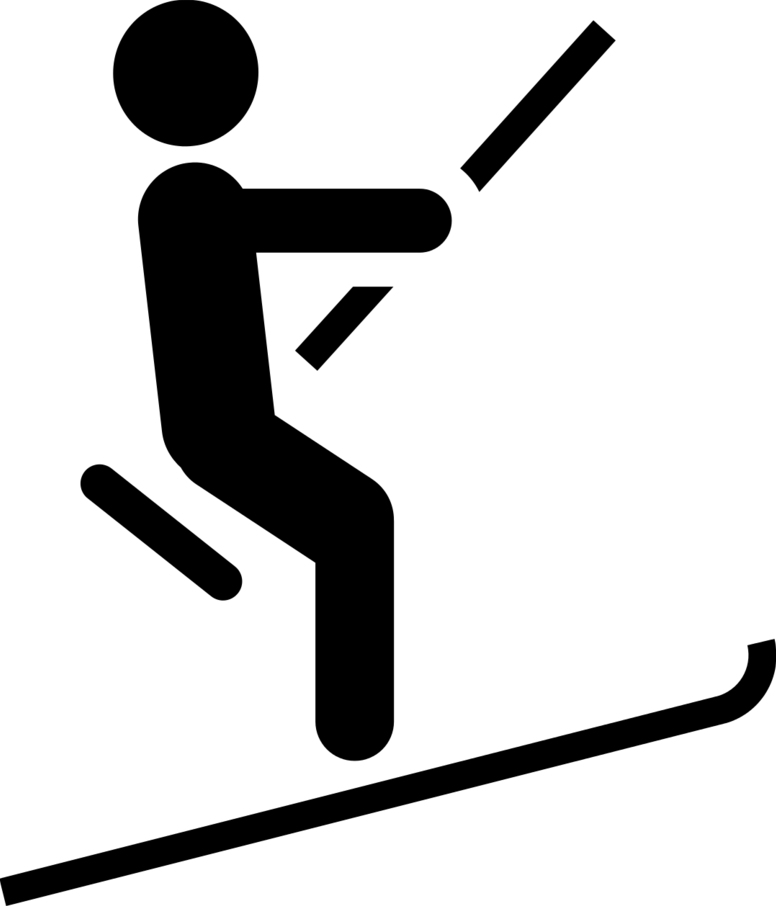}%
  \endgroup
}
\newcommand{\chairLift}{%
  \begingroup\normalfont
  \includegraphics[height=1.3\fontcharht\font`\B]{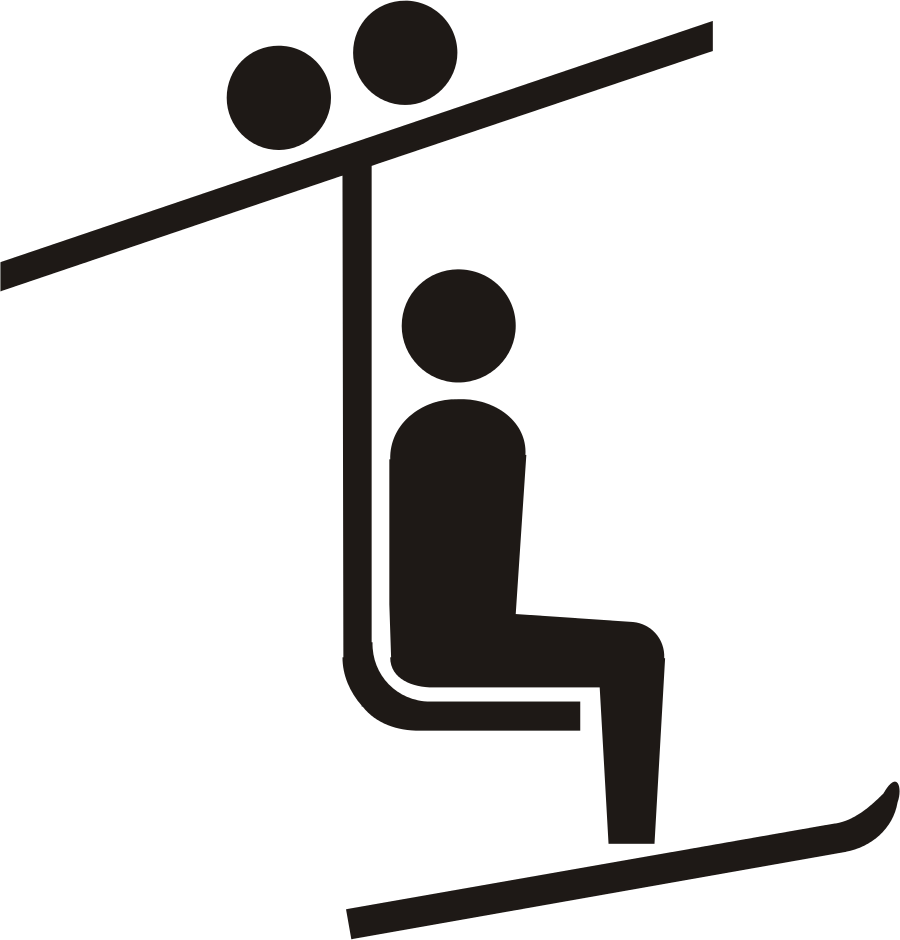}%
  \endgroup
}
\newcommand{\gondolaLift}{%
  \begingroup\normalfont
  \includegraphics[height=1.3\fontcharht\font`\B]{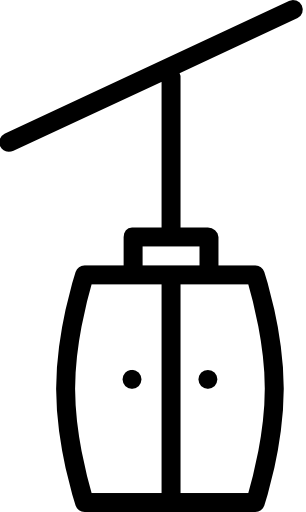}%
  \endgroup
}
\newcommand{\cableCarLift}{%
  \begingroup\normalfont
  \includegraphics[height=\fontcharht\font`\B]{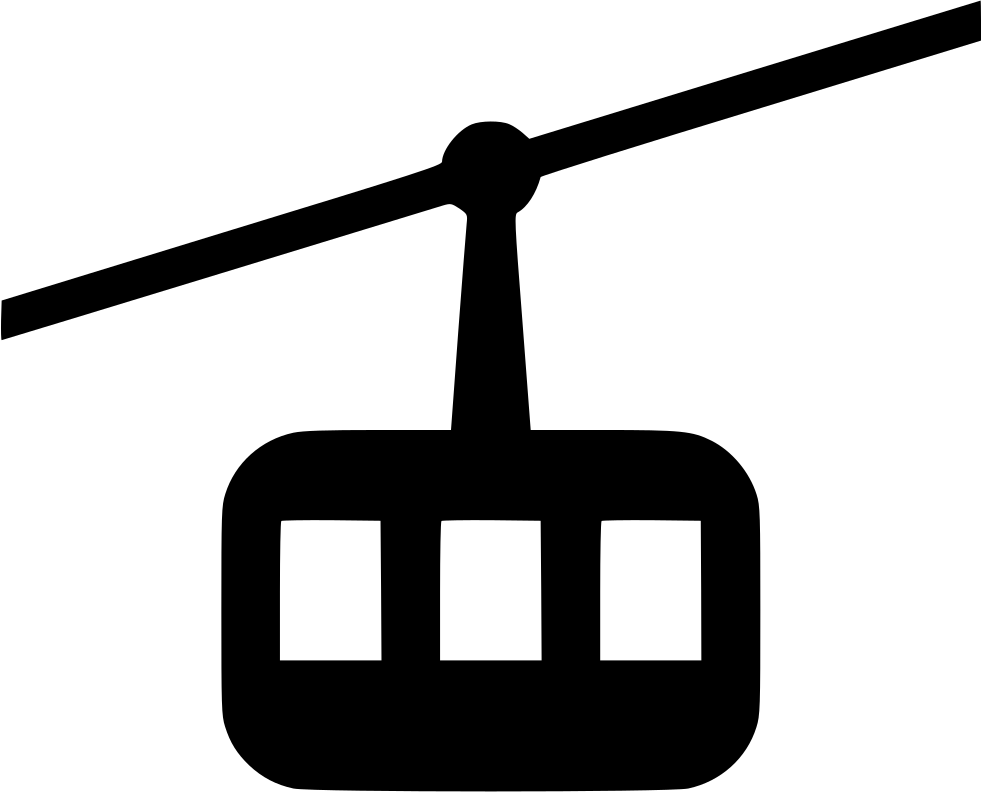}%
  \endgroup
}
\newcommand{\skierIcon}{%
  \begingroup\normalfont
  \includegraphics[height=\fontcharht\font`\B]{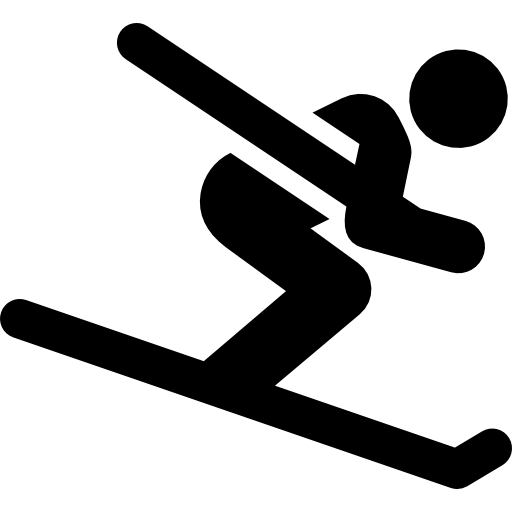}%
  \endgroup
}

\newcommand{\boxSep}{0pt}
\newcommand{\boxRule}{0.5pt}
\newcommand{\boxTopLine}{2pt}
\newcommand{\boxBottomLine}{0pt}
\newcommand{\boxLeftLine}{1.2pt}
\newcommand{\boxRightLine}{1.2pt}
\newcommand{\boxArc}{2pt}

\newcommand{\appurl}{https://skirouting.dbvis.de}
\newcommand{\skivis}{\href{\appurl}{SkiVis}}

\definecolor{c_comment}{HTML}{008000}
\definecolor{c_comment}{HTML}{000000} 
\newcommand{\jr}[1]{{\textcolor{c_comment}{#1}}}

\definecolor{c_vis_rev}{HTML}{EE0000}
\definecolor{c_vis_rev}{HTML}{000000} 
\newcommand{\visRev}[1]{{\textcolor{c_vis_rev}{#1}}}

\definecolor{todo_comment}{HTML}{FF0000}

\definecolor{c_miller}{HTML}{dd22dd}

\definecolor{rbuch}{HTML}{6f11bf}

\newcommand{\guideA}{\p{G}{1}} 
\newcommand{\guideB}{\p{G}{2}} 
\newcommand{\guideC}{\p{G}{3}} 
\newcommand{\skierA}{\p{P}{1}} 
\newcommand{\skierB}{\p{P}{2}} 
\newcommand{\skierC}{\p{P}{3}} 

\newcommand{\p}[2]{{$\mathbf{#1}_\mathbf{#2}$}}

\definecolor{label_a}{HTML}{8282ff}
\definecolor{label_b}{HTML}{ff8282}
\definecolor{label_c}{HTML}{828282}
\definecolor{label_d}{HTML}{ffcd82}
\definecolor{label_e}{HTML}{ffff82}
\definecolor{label_f}{HTML}{82ffa3}

\newcommand{\taskA}{\textbf{\texttt{[T1]}}}
\newcommand{\taskB}{\textbf{\texttt{[T2]}}}
\newcommand{\taskC}{\textbf{\texttt{[T3]}}}
\newcommand{\taskD}{\textbf{\texttt{[T4]}}}

\newtcbox{\rectLabelA}{enhanced,nobeforeafter,tcbox raise base,boxrule=\boxRule,top=\boxTopLine,bottom=\boxBottomLine,
  right=\boxRightLine,left=\boxLeftLine,arc=\boxArc,boxsep=\boxSep,before upper={\vphantom{dlg}},
  colframe=label_a,coltext=white,colback=label_a}
\robustify{\rectLabelA}
\newcommand{\aLabel}{\rectLabelA{A}}

\newtcbox{\rectLabelB}{enhanced,nobeforeafter,tcbox raise base,boxrule=\boxRule,top=\boxTopLine,bottom=\boxBottomLine,
  right=\boxRightLine,left=\boxLeftLine,arc=\boxArc,boxsep=\boxSep,before upper={\vphantom{dlg}},
  colframe=label_b,coltext=white,colback=label_b}
\robustify{\rectLabelB}
\newcommand{\bLabel}{\rectLabelB{B}}

\newtcbox{\rectLabelC}{enhanced,nobeforeafter,tcbox raise base,boxrule=\boxRule,top=\boxTopLine,bottom=\boxBottomLine,
  right=\boxRightLine,left=\boxLeftLine,arc=\boxArc,boxsep=\boxSep,before upper={\vphantom{dlg}},
  colframe=label_c,coltext=white,colback=label_c}
\robustify{\rectLabelC}
\newcommand{\cLabel}{\rectLabelC{C}}

\newtcbox{\rectLabelD}{enhanced,nobeforeafter,tcbox raise base,boxrule=\boxRule,top=\boxTopLine,bottom=\boxBottomLine,
  right=\boxRightLine,left=\boxLeftLine,arc=\boxArc,boxsep=\boxSep,before upper={\vphantom{dlg}},
  colframe=label_d,coltext=black,colback=label_d}
\robustify{\rectLabelD}
\newcommand{\dLabel}{\rectLabelD{D}}

\newtcbox{\rectLabelE}{enhanced,nobeforeafter,tcbox raise base,boxrule=\boxRule,top=\boxTopLine,bottom=\boxBottomLine,
  right=\boxRightLine,left=\boxLeftLine,arc=\boxArc,boxsep=\boxSep,before upper={\vphantom{dlg}},
  colframe=label_e,coltext=black,colback=label_e}
\robustify{\rectLabelE}
\newtcbox{\rectLabelF}{enhanced,nobeforeafter,tcbox raise base,boxrule=\boxRule,top=\boxTopLine,bottom=\boxBottomLine,
  right=\boxRightLine,left=\boxLeftLine,arc=\boxArc,boxsep=\boxSep,before upper={\vphantom{dlg}},
  colframe=label_f,coltext=black,colback=label_f}
\robustify{\rectLabelF}
\teaser{
\href{\appurl}{\includegraphics[width=\linewidth]{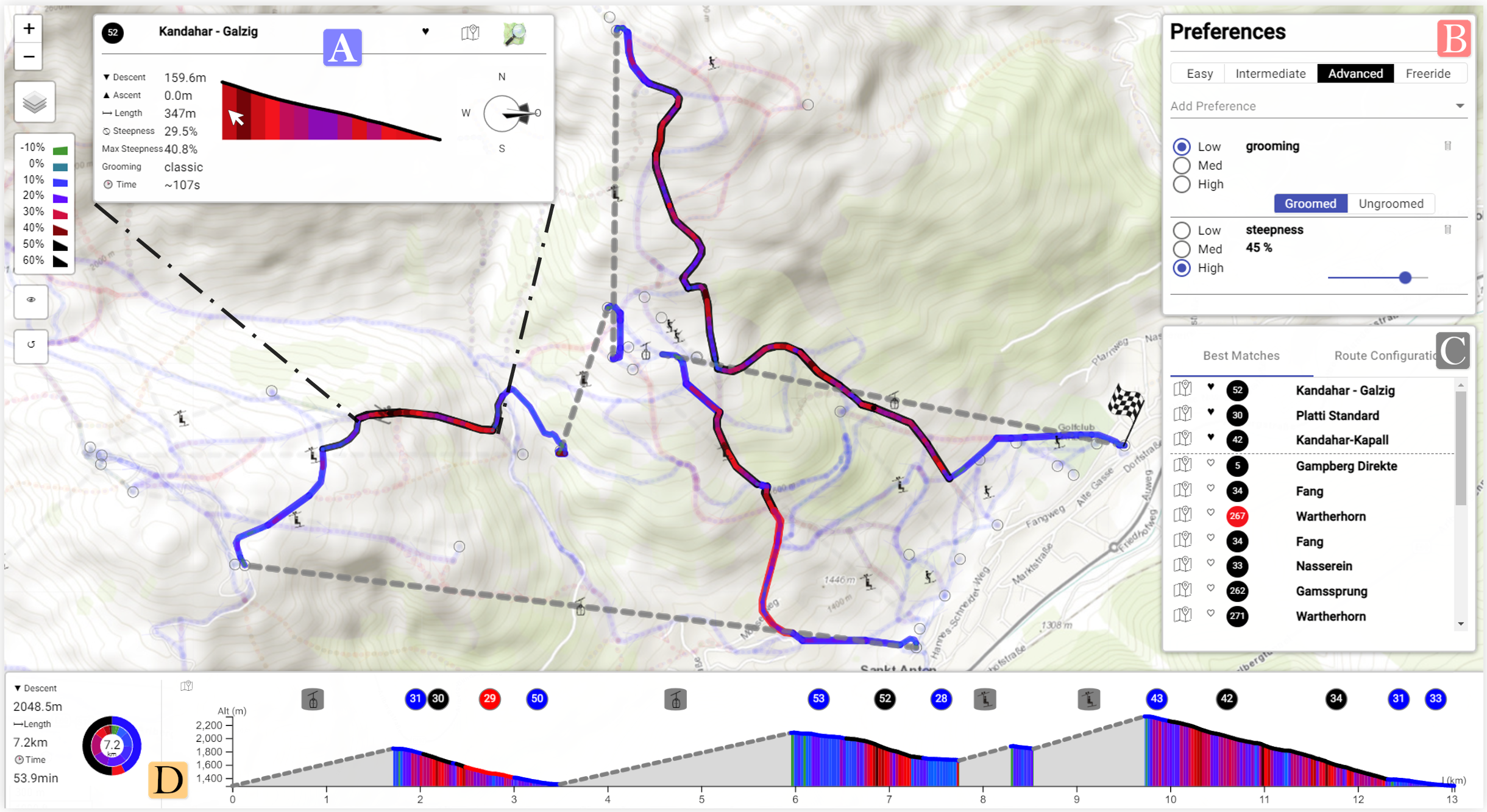}} 
 \centering
  \caption{Our application interface visualizes the slopes and lifts of a ski resort on a topographic map, allowing a detailed exploration through multiple linked and brushed components~\aLabel. 
  By applying preferences~\bLabel, slope recommendations are provided~\cLabel~that can be integrated into a route planning algorithm.
  The generated route can be investigated further regarding its sequence and steepness profile in the \textit{route summary}~\dLabel.
  }
\label{fig:teaser}
}

\graphicspath{{figs/}{figures/}{pictures/}{images/}{./}} 

\usepackage{mathptmx}                  

\begin{document}



\maketitle

\section{Introduction}

\label{section:intro}




Outdoor activities are prominent in many fields, including education~\cite{ijerph18052506}, entertainment~\cite{STIGSDOTTER2011295}, health~\cite{mccurdy2010using}, and team-building\cite{owen_renee}. 
Especially mountain regions are favored for outdoor activities due to scenic outlooks, fresh air, and diverse flora and fauna~\cite{carden2017simply}.
There are various mountain outdoor activities such as hiking, cycling, and skiing. 
Every year, approximately 400 million skiers find their way into ski resorts worldwide~\cite{skitourismreport:vanat:21}.
Larger ski resorts tend to be more profitable~\cite{profitinskiresorts:falk:20}, leading to an increased number of linked lift constructions connecting neighboring resorts~\cite{resortlinking:FALK:17}.
The resulting joined areas become even more complex and hence are more difficult to navigate, enhancing the need for routing support. 
Since the number of possible slope route combinations is increasing, skiers and snowboarders face the challenge of keeping a good overview.
Similar outdoor activities to skiing, such as hiking~\cite{tirla2014digital,lera2017analysing} or bicycling~\cite{vancouvercycle:su:10, multirouteplanning:zhu:20}, provide solutions for orientation and route planning. 
While some concepts are transferable, skiing trip navigation poses specific challenges, such as high traffic on a comparatively smaller route network, a frequent alteration between the use of slopes and lifts, and the fact that slopes can only be traveled in one direction.
In addition, the demand for risk management, such as avoiding causes for avalanches, is likewise increasing~\cite{sterchi2019avalanche}, especially for slopes that invite to run off-piste with snow powder conditions.
\clearpage
\noindent Weather events, especially snow conditions and sun hours, can also have a considerable impact on the number of visitors to ski resorts.
On sunny days with plenty of snow, the slopes and restaurants are typically crowded, resulting in disliked waiting times at lift stations~\cite{weatherinresorts:haugom:19}.
Conversely, bad weather negatively influences customer demand~\cite{weatherpricing:Malasevska:20} and ticket sales~\cite{weatherimpact:shih:09}.
A lack of snow can either be mitigated by using unfavorable artificial snow~\cite{realsnow:picketing:10} or cause certain slopes in lower altitudes to remain closed.
Heavy winds or thunderstorms can force the resort to shut down entirely.
While ski resorts have been researched concerning climate change~\cite{climatechange}, economic sustainability~\cite{ecosus:moreno:18}, demand prediction~\cite{skidaysprediction}, customer analysis \cite{skiresortvisitors}, or pricing models~\cite{resortprices:haugom:21, dynamicpricing:alnes:21, weatherpricing:Malasevska:20}, only a few research projects focus on navigating along ski slopes.

\begin{figure}[t]
    \centering
    \href{https://www.skiarlberg.at/de/Ski-Arlberg/Live-Infos/Interaktive-Karte}{\includegraphics[width=\linewidth]{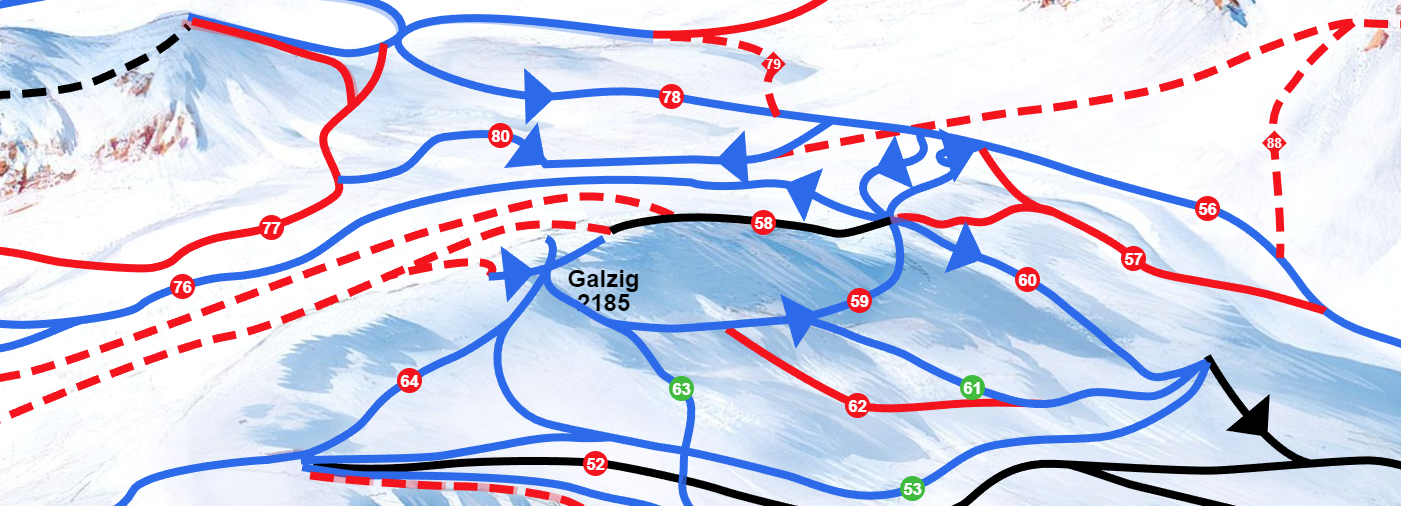}}    
    \caption{Excerpt of the interactive Ski Arlberg slope map~\cite{arlbergmap}.}
    \label{fig:slopemap}
    \vspace{-1.5em}
\end{figure}

The main resource for navigation in ski resorts is a ski slope map, as seen in~\autoref{fig:slopemap}, which is designed uniquely for every resort.
%
Instead of depicting the area in a topographically accurate manner, most ski resorts employ a more aesthetically pleasing and panoramic style to artistically illustrate the landscape~\cite{skimaps}. 
\jr{While this partly marketing-focused design compactly depicts the entire resort}, it can compromise routing capabilities and thereby lead to skiers' misorientation~\cite{deformedmap}.
%
From a graph theory viewpoint~\cite{graphtheory:riaz:11}, ski resorts can be modeled as a geospatial, partly directed network consisting of downhill slopes and uphill lifts as edges and intersections as nodes. 
Occasionally, some lifts bi-directionally link two otherwise separated ski resorts.
There are several types of lifts: \textit{t-bar}~\tBarLift, \textit{chair lift}~\chairLift, \textit{gondola}~\gondolaLift, and \textit{cable car}~\cableCarLift, depending on the skiing resorts' size and natural conditions. 
Slopes are usually classified according to their difficulty and labeled by a color-coded system, e.g., \textit{easy}~\bluecircle, \textit{intermediate}~\redcircle, and \textit{advanced}~\blackcircle~in Austria~\cite{onorm} or Switzerland. 
Depending on the country where a ski resort is located, slopes with similar characteristics are labeled using different difficulty classification systems~\cite{pisteprofile:bortnyk:20}. 
Such differences can lead to confusion among visitors who are typically used to a single classification strategy.
Due to the simplicity of the color classification, hazards such as narrow slopes or annoying uphill segments can not be identified with the current design.

The steepness serves as a general indication for difficulty~\cite{onorm}, with steeper slopes being ranked as more challenging.
However, the difficulty label is often imprecise since solely the steepest subsegment is typically responsible for classifying the entire slope. 
Therefore, it is reasonable to consider other, more detailed characteristics of slopes, such as the steepness or compass direction of all its subsegments.
The compass direction influences the amount of solar radiation a slope is exposed to. 
In the northern hemisphere, north-facing slopes experience more day hours of direct sun radiation, thus exhibiting different snow conditions. 
At extremely low temperatures, northern slopes are often frozen and more difficult to ski, especially for beginners. 
In contrast, when temperatures rise at the end of a skiing season, southern slopes become muddy and unattractive, especially for skiers with aging ski gear.
In such situations, it is advisable to use southern slopes in the morning and switch to slopes that are not directly facing the sun later.

\jr{ 
Skiing enthusiasts, ranging from intermediate or advanced skiers to ski guides, can greatly benefit from domain-specific visual analytics solutions.
Offering customized visualizations can provide them with a novel viewpoint on slopes within a ski resort, leading to insights they would not have otherwise. 
Especially when traveling to previously unknown resorts, interactive route planning capabilities can support skiers in identifying their preferred slopes and suggest route recommendations according to their individual preferences.  
}
Besides skiers, other ski resort visitors, such as snowboarders and sledgers, exhibit different characteristics~\cite{skierprefs:Leutschner:71}. 
Addressing the needs of each visitor group requires an adaptive, user-centered methodology.
In some ski resorts, downhill valley runs are shared by both sledgers and skiers, which requires all slope users to pay extra attention. 
Otherwise, it could lead to severe downhill slope accidents, which do not seldom occur~\cite{chaze2008headinjuries}.

The limitations of current approaches in the ski domain led us to investigate the following research questions:
\begin{itemize}
    \item \textit{How can we visualize a ski resort as a geospatial network while effectively incorporating individual slope properties? }
    \item \textit{How can we facilitate preference-based routing capabilities within a ski resort?}
\end{itemize}

\subhead{ Contributions --} 
This paper contributes a requirement and task analysis for modeling a ski resort depending on the underlying properties of its lifts and slopes. 
In collaboration with domain experts, we formalize the issues at hand and provide a list of task characteristics.
A major contribution is our user-centered Visual Analysis prototype to explore a ski resort slope network. 
A central part of the prototype enables the user to interactively receive route recommendations tailored to slope preferences.
Further, we conducted a qualitative evaluation to assess the applicability of our approach and discuss its benefits and shortcomings.
Finally, we present open research opportunities to encourage further research within the domain of visual analytics for ski resorts.

\section{Related Work}
Since the research conducted in this paper covers the domains of ski visualizations and routing in geospatial networks, we dedicate a section for discussing related approaches to each of these topics.

\subhead{Ski Visualizations --}
A survey on ski trail maps in North America~\cite{skimaps} determined a dominance of the painted panorama style (see~\autoref{fig:slopemap}) (86\%), which distorts the underlying topography to emphasize certain memorable features. 
According to Balzarini et al.~\cite{deformedmap}, this so-called \textit{terrain invention} can introduce difficulties in interpretation, such as misunderstanding or uncertainty.
Field~\cite{skimetromap} investigated the effectiveness of a schematic metro map in contrast to the dominating panorama-styled maps and proved its usefulness for solving on-mountain navigation tasks.
However, the absence of geographical features relevant for orientation and the familiarity with traditional panorama maps determined no clear advantage of the proposed layout, especially for beginners.
For additional information on schematic maps in general, we refer to a survey of Wu et al.~\cite{metromapsurvey:wu:20}. 
To enable preference-based slope run selection, Dunlop et al.~\cite{pistepreferences} conducted experiments with varying the line width in slope maps according to a user's preferences. 
Their evaluation showed that the time required for route planning can be reduced when implementing preference-based visualization techniques. 
Regarding slope properties, Bortnyk et al.~\cite{pisteprofile:bortnyk:20} investigated a Ukrainian ski resort based on the slopes' steepness. 
All of the above-mentioned approaches adhere to the standard of visualizing slopes by a color-coded line according to their difficulty. 
Thus, they fail to address the diversity of different slopes regarding their geographical properties.

Web-based, interactive systems such as OpenSkiMap~\cite{openskimap} or FATMAP~\cite{fatmap} display further information on individual slopes besides steepness, such as distance, altitude, and compass direction, but do not comprise preference-based route planning features.
\jr{
Other mobile applications within the ski domain offer progress tracking~\cite{strava, slopesapp}, or details on snow conditions~\cite{opensnow, onthesnow}, but also lack routing features and are limited regarding additional slope information.
}
Friedsam et al.~\cite{osmskirouting} provide routing capabilities based on \textit{OpenStreetMap} (OSM) data, however, they do not take user preferences into account.
Augmented Reality systems provide the skier with visual and auditory signals at junctions or safety-critical areas~\cite{ridersguide:schmid:11} or enable collaboration between skiers to share personalized information on a digital trail map~\cite{skiar:fedosov:16}.
Aside from the confined area of controlled ski resorts, the mitigation of avalanche risks has received attention from researchers, using GPS tracking data to investigate the behavior of out-of-bounds skiers~\cite{outofboundsski} or explore the terrain preferences of heli-ski guides~\cite{heliski:HENDRIKX:16}.

In the closely related hiking domain, Bleisch and Dykes~\cite{virtualhikeplan:bleisch:08} conducted a study to evaluate the usefulness of 3D visualizations, which concludes that 3D layouts outperform 2D maps for general overview tasks but not for route detection and planning.
Human mobility patterns of hiking activities are investigated by Lera et al.~\cite{lera2017analysing}.
Tîrlă et al.~\cite{tirla2014digital} visually compared elevation profiles of different hiking trails, and
Korohoda et al.~\cite{hikingobstacles:korohoda:21} leveraged volunteered geographic information (VGI) and digital elevation model (DEM) information to assess the difficulty classification of hiking routes and to detect obstacles.



\subhead{Routing in Geospatial Networks --}
The idea of routing in geospatial networks has been investigated from several subfields of information science, such as routing algorithms, decision-making, and route visualizations. 
Felício et al.~\cite{osmrouting:felicio:2022} describe how to leverage OSM data and integrate information into a multi-criteria route planning approach.  
A similar approach by Guo et al.~\cite{preferencebasedvehiclerouting} facilitates vehicle routing by adapting the edge costs of a network structure according to a given context and preferences, providing personalized route recommendations.

Various approaches to the shortest path problem, many of which factor in multiple criteria, have been formulated in the domain of cycling. 
Storandt~\cite{storandt_2012} considers a hard constraint on the height difference to minimize energy consumption along the path.
Hrncir et al.~\cite{multibicyclerouting} optimize concerning the travel time, comfort, and minimum elevation gain, Song et al.~\cite{paretocycling} efficiently compute full Pareto sets for these criteria. 
Zhu~\cite{multirouteplanning:zhu:20} further computes a tourist satisfaction score as a criterion that influences the optimal route. 
The reasoning and driving factors behind routing choices from trajectory data of bike-sharing services have been studied by Scott et al.~\cite{bikeprefs:scott:21}. 
Molokac et al.~\cite{hikingchoices:molokac:22} used survey data to analyze the preferences regarding hiking trail choices.

Regarding route visualizations, Haunert and Sering~\cite{roadnetworkfocusregion} propose the usage of focus regions, where the area of interest is scaled up for better visibility.
Wu et al.~\cite{tripcenteredmetrolayout} introduce the concept of travel-route-centered maps, a custom metro map layout where the map is rotated and adapted to visualize the route path in a straight line.
\visRev{In the context of tourism trip planning, Zhang et al.~\cite{triplan:Zhang:22} and Deng et al.~\cite{tourplanning:deng:2022} propose a visual analytics framework using multiple linked views to facilitate interactive tour planning based on popular subtours from crowdsourced data.}
Su et al.\cite{vancouvercycle:su:10} provide a route planning interface for cycling in Vancouver, incorporating user preferences regarding steepness, air pollution, or vegetation.
\visRev{The domain of urban visual analytics~\cite{urbanva:Deng:2023}, particularly the optimization of bus routes~\cite{allaboard:DiLorenzo:16, shuttlebus:Liu:9331263, betterbus:weng:21}, further relates to this work.}

\subhead{Research Gap --}
While several approaches for route planning focus on other outdoor activities such as cycling network analysis~\cite{storandt_2012, multibicyclerouting, bikeprefs:scott:21} or hiking trails\cite{tirla2014digital}, only little research exists for the ski domain~\cite{osmskirouting}.
Despite its discussed importance, current systems neglect user-specific preferences.
Moreover, established solutions for exploring ski resorts~\cite{openskimap, fatmap} lack customization and do not consider human-in-the-loop concepts for knowledge generation~\cite{knowledgegeneration:sacha:14}.
To the best of our knowledge, for the domain of ski resorts, no analysis approach exists that enables the visual exploration of the resort and interactively provides routing recommendations based on user-defined preferences.









\section{Requirement Analysis}

\label{section:requirements}
To precisely formulate the requirements for a workspace that addresses visual exploration and route planning in ski resorts, we consulted with ski guides and expert skiers about their opinions towards this initiative.
\jr{In the context of a pre-study, we had informal discussions with four experts \visRev{(>15 years of yearly skiing experience, two being certified ski guides)} to learn about the currently available tools, the limitations they encounter while using them, and their expectations towards novel software.}
\jr{The most commonly utilized tools~\cite{fatmap, arlbergmap, bergfex} do not provide route planning functionalities and only provide limited contextual information about individual slopes.}
Based on the experts' feedback, the following section frames the abstract model of the workspace, with special regard to the utilized data and supported tasks.
The target audience of the system comprises ski guides as well as intermediate and advanced skiers.
While novice skiers are cognitively occupied with acquiring various skiing techniques, routing is not their primary concern, and typically they travel with a group where a more experienced skier is responsible for the route planning. 
The system is designed to provide the best results when guiding the user in unfamiliar resorts. However, gaining insights and optimal route recommendations from resorts with which the user is already familiar with is an equally important objective.

\subhead{Feature Characteristics --}
The sport of Skiing is closely coupled to the environment of mountain regions. 
Thus, it is apparent that geographic features can be used to model the geospatial network of slopes and lifts. 
According to domain experts, the \textit{steepness} of a slope is the most significant feature and serves as the main criteria for determining the difficulty of a slope in Austria \cite{onorm}. 
Another important feature is the \textit{compass direction}. 
In the northern hemisphere, north-facing slopes are less exposed to sunlight, resulting in different snow conditions. 
Further slope features to note are the \textit{length}, \textit{altitude}, \textit{width}, or \textit{curvature}. 
Very narrow or extremely curvy slopes can impose a challenge, especially for novice skiers.
Aside from geographical features, further attributes can also be considered important for the analysis of ski resorts. Especially in peak season, slopes and lifts can become \textit{crowded}, resulting in undesired waiting times. The majority of slopes in a resort are \textit{groomed} daily by means of a snowcat. However, some slopes remain purposely \textit{ungroomed} and require a different skiing technique. 
Dynamic properties further contribute to the model, albeit being more complicated to record and forecast. These include meteorologic conditions such as \textit{temperature}, \textit{wind}, \textit{fog}, \textit{snowfall}, or \textit{snow conditions} in general.
\jr{Since accurate readings for these properties are unobtainable on a slope-level resolution, their impact on deciding which route to take is arguably limited. 
Therefore, we decided to exclude these features in the development of our initial prototype.}

\subhead{Task Requirements --}
As already outlined in the previous passage, the edges of the geospatial network of a ski resort can be defined by a set of different features. 
The exploration of these features (i.e., what is the steepness of a certain segment of a given slope) constitutes a pivotal objective~\taskA.
Since skiers have individual preferences regarding these features, a ranking of slopes according to their given preferences should be incorporated to identify favorite slopes~\taskB.
Concerning routing, the system should be able to provide routing recommendations between any arbitrary points in the network, including the possibility of a round trip, as skiers usually reside in a lodge or hotel and finish their daily route at the exact same location where it started~\taskC.
Following~\taskB, \visRev{the routing recommendations should be optimized to reflect the personal user preferences along the entire route}~\taskD. 
To summarize, in collaboration with domain experts, we elicit the following four tasks for our further analysis:

\vspace{.5em}

\subheadnvs{\taskA}~Exploration of a ski resort according to the defined features \\
\subheadnvs{\taskB}~Provide a ranking based on preferences toward these features \\
\subheadnvs{\taskC}~Allow routing between two arbitrary network points \\
\subheadnvs{\taskD}~Integrate preferences into the routing algorithm \\


\section{Application Design}
\label{section:application}
This section introduces the developed workspace, the used data, the required preprocessing steps, design choices regarding the individual components, and recommended interaction workflows.
\skivis~can be accessed under the following domain: \href{\appurl}{\appurl}.

\subhead{Ski Arlberg --} Labeled as the \textit{cradle of alpine skiing}, the area of Ski Arlberg is the largest connected ski resort in Austria, consisting of 88 lifts and a total slope length of 302 km~\cite{arlberg}. It connects the three regions of St. Anton - St. Christoph - Stuben, Lech - Oberlech - Zürs, and Warth-Schröcken with a series of connecting lifts, resulting in a large, complex network structure. 
\jr{According to~\cite{skitourismreport:vanat:21}, it is the resort with the second most annual skier visits worldwide between 2016 and 2021.}
While the techniques applied in this paper are applicable to virtually any other ski resort, we chose Ski Arlberg due to its size and complexity as the most appropriate region to showcase the potential.  

\subsection{Ski-Resort Related Data}
In the subsequent section, we discuss the data sets used to model the feature characteristics introduced in~\autoref{section:requirements}, as well as preprocessing methods that were implemented to enhance the data quality. 

\subhead{Data Sources --} 
\jr{
Detailed geographic and contextual information about individual slopes and lifts is challenging to acquire. 
While most resorts provide an interactive, online version of their slope map containing a list of the slopes and lifts~\cite{arlbergmap}, no geographic information about the length, width, or actual trajectory is present. 
}
We opted to obtain this information from OSM~\cite{osm:osm:17}, a community-owned, collaborative database for geographic data. 
Depending on the ski resort, additional information such as the name of the slope, its difficulty classification, the reference number, or grooming conditions is available aside from the spatial line component.
To attain details about travel durations or the popularity of slopes, analyzing human mobility data can be insightful.
We opted for using VGI data from \textit{Strava}~\cite{strava}, where users can record and share their exercise data of different leisure activities, such as skiing. 
We extracted trajectory data recorded on the slopes of the Ski Arlberg resort \jr{between 2008 and 2022}, resulting in more than 15.000 activities.
These trajectories serve as the basis for our calculations to gain insights on the crowdedness, as well as an estimate for the required skiing time for each slope \jr{and lift alike}.



\begin{figure}[t!]
    \centering    
     \includegraphics[width=\linewidth]{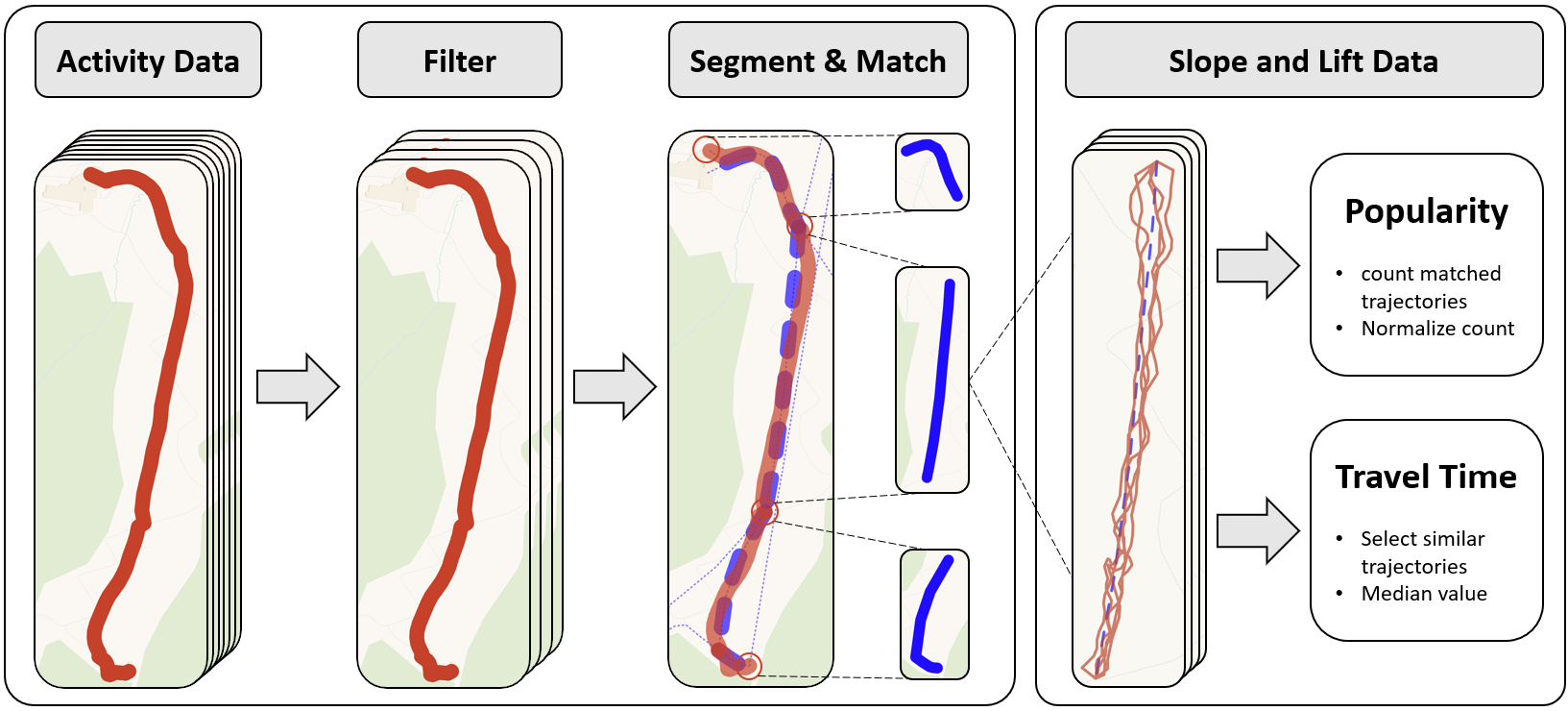}
     
    \caption{Preprocessing pipeline.}
    \label{fig:preprocessing}
    \vspace{-1.5em}
\end{figure}



\subhead{Data Preprocessing --}
Since the OSM \jr{and the extracted trajectory data} are community-maintained data sources, a data preprocessing step is indispensable to ensure satisfying data quality. 
\jr{
To facilitate a comparison between slopes and enable routing based on preferences, we need to 
construct a connected network structure from our slope dataset and derive a value for every feature on a fine-grained spatial resolution.
By means of the \textit{pgRouting} package~\cite{pgrouting}, a geospatial line-based dataset can be extended with routing functionalities.}
In our initial exploration of the network topology created by \textit{pgRouting}, we noticed connectivity issues at some network nodes, where the slope endings were not connected to the closest lift vertex. 
\jr{ A total of 94 dead-end slopes were detected in the topology.}
To improve the connectedness of the routing network, all vertices with a degree of 1 (nodes with only one connection) were checked for close, unconnected neighbor vertices within a spatial proximity of 30 meters.
If such a close neighbor was found in the dataset, a helper slope was inserted to connect these vertices. 
80 such helper connections were additionally inserted into the database, \jr{reducing the number of dead-end slopes to 15.}

For a more granular analysis of the slope network, each slope was divided into smaller equidistant subsegments, ~\jr{where a value of 30 meters of length provided the best results concerning the accuracy, visual clutter, and performance.}
The DEM from the Copernicus earth observation programme~\cite{copernicus} was utilized to determine the \textit{altitude}, \textit{steepness}, and \textit{compass direction} of every subsegment independently to enable a more precise analysis.

\jr{Details regarding the \textit{popularity} and \textit{travel duration} were extracted from the GPS trajectories according to the preprocessing pipeline in~\autoref{fig:preprocessing}.
As GPS tracking can provide infrequent temporal resolution results in mountain regions, we filtered out 1874 activities where the median time between subsequent timestamps was greater than one second.
Since the activity data is typically recorded over an entire day, the trajectory must be segmented into sub-trajectories corresponding to the ski slopes and lifts.
We followed existing trajectory segmentation approaches~\cite{mapmatching:Jensen:2009, neat:han:12}, which employ the concept of map-matching when road network context is available.
To introduce additional robustness, we exploited the difference in altitude to segment the activity into lift rides (increasing altitude) and ski rides (decreasing altitude) prior to the map-matching procedure.
Afterward, we split the trajectories on intersection points into our desired sub-trajectories, each corresponding to a slope or lift.
We extracted a total of 1.8 million such sub-trajectories, with 1088 entries on average per slope.}

Due to privacy reasons, the obtained GPS trajectory data contains only temporal information relative to the first datapoint of the series, concealing the precise time of day when the activity commenced. 
This merely permits a coarse estimation about \textit{popularity} and \textit{average travel duration} of each slope and rules out a more detailed dynamic traffic analysis of the entire resort.
\jr{By counting the number of matched trajectories for every slope, we can determine a popularity ranking, which we obtained by logarithmic normalization to reduce skewness in the output value.
In principle, the mean travel time can be evaluated by calculating the median between the difference of the start and end timestamp of the trajectories associated with each slope or lift.
To prune against outlier, we consider only the trajectories most similar to the initial slope.
For slopes with insufficient trajectory records (<10\% of the average per slope (1088) = 108) available, we had to interpolate the travel time from subsegments with a similar steepness value.
}



\begin{figure}[t!]
     \centering
     \begin{subfigure}[b]{0.174\linewidth}
         \centering
         \includegraphics[width=\textwidth]{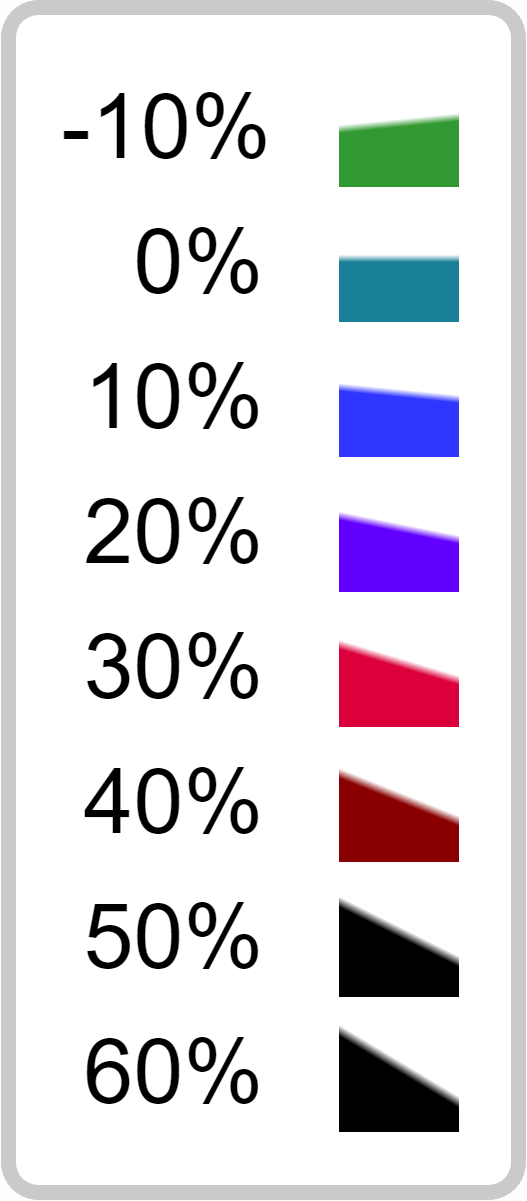}
         \caption{}
         \label{fig:colorlegend}
     \end{subfigure}
     \hfill
     \begin{subfigure}[b]{0.18\linewidth}
         \centering
         \includegraphics[width=\textwidth]{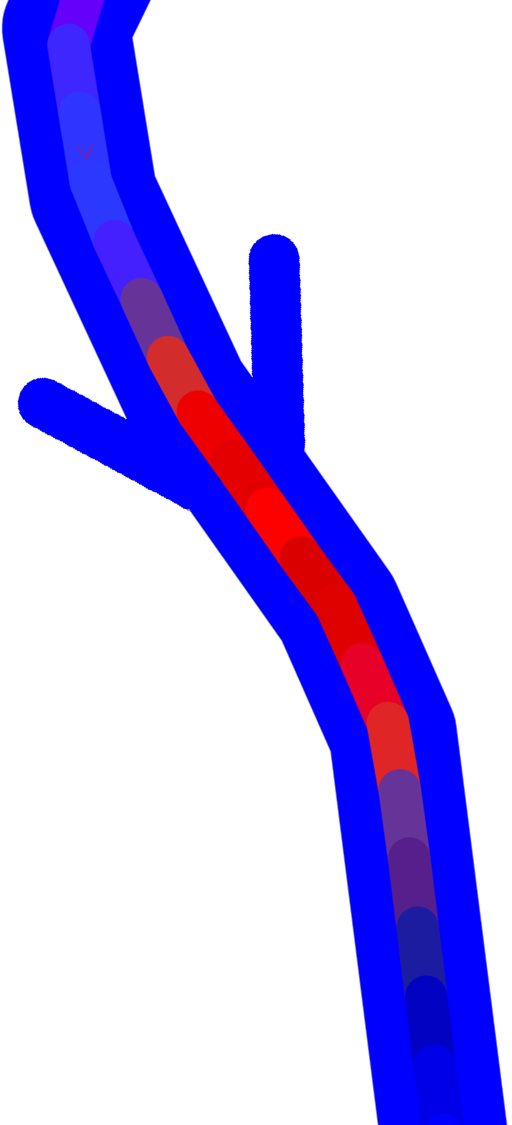}
         \caption{}
         \label{fig:slopea}
     \end{subfigure}
     \hfill
     \begin{subfigure}[b]{0.18\linewidth}
         \centering
         \includegraphics[width=\textwidth]{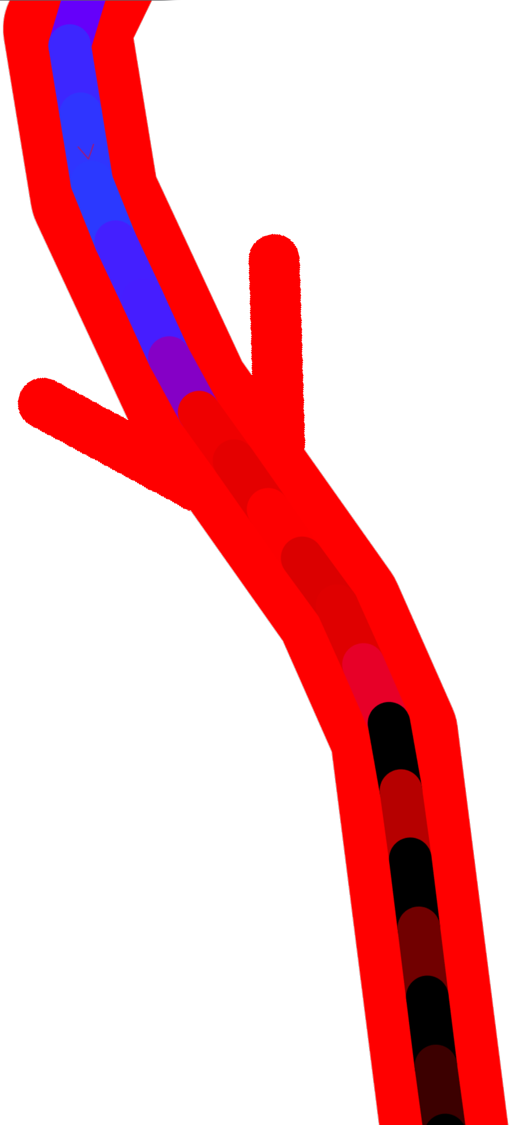}
         \caption{}
         \label{fig:slopeb}
     \end{subfigure}
     \hfill
     \begin{subfigure}[b]{0.18\linewidth}
         \centering
         \includegraphics[width=\textwidth]{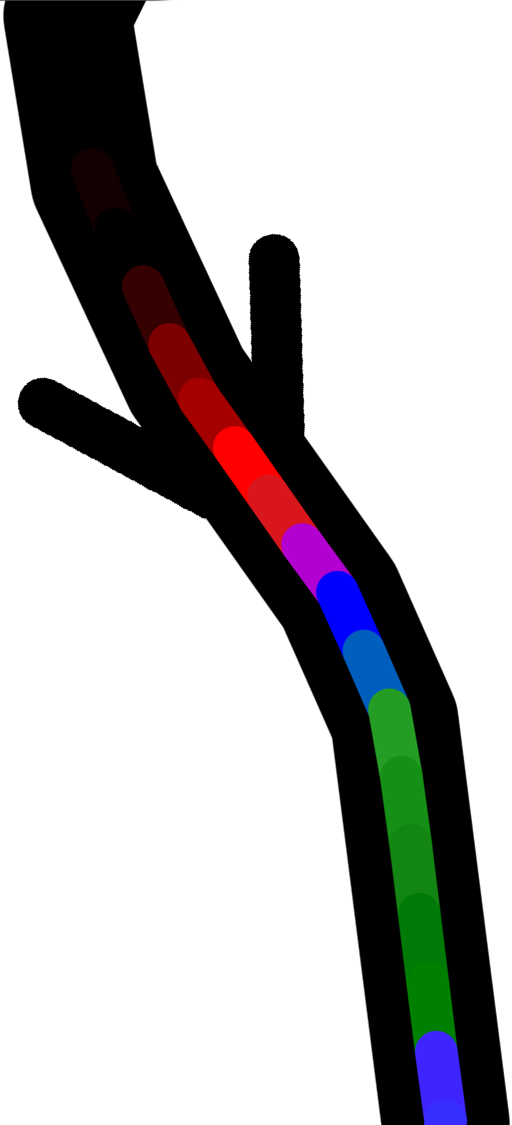}
         \caption{}
         \label{fig:slopec}
     \end{subfigure}
     \hfill
     \begin{subfigure}[b]{0.175\linewidth}
         \centering
         \includegraphics[width=\textwidth]{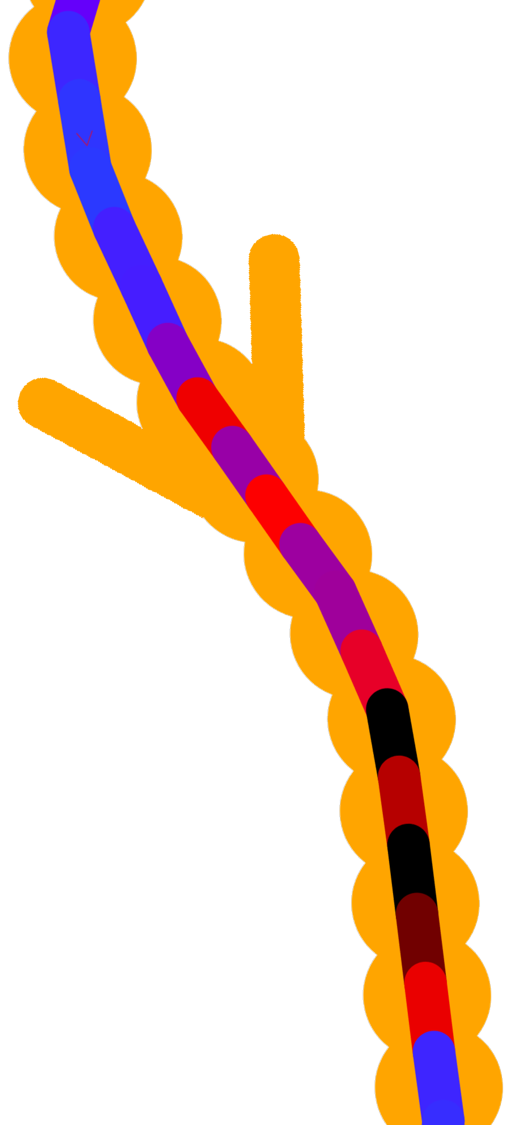}
         \caption{}
         \label{fig:sloped}
     \end{subfigure}
     \vspace{-1em}
        \caption{Visual design of slopes. The outer line of slopes displays the color-coded difficulty established by the resort. The inner line encodes the actual steepness along the slope using our colorscale (a) and exposes patterns such as steep passages on easy slopes (b), intermediate slopes with varying steepness profile (c), uphill segments on advanced slopes (d), or ungroomed \visRev{(shown with a dotted line)} freeride slopes (e).}
        \label{fig:designspace}
        \vspace{-1.5em}
\end{figure}
\subsection{Design Rationale}
This section outlines design choices of the workspace, in particular the map visualization, the routing algorithm, and the interaction workflows.

\subhead{Map Visualization --}
The application interface consists of a topographical 2D map as its core component. The visualization aims to encode as many features as possible~\taskA~without overextending the cognitive load of the user. In contrast to the vast majority of ski trail maps, we decided against a panoramic layout in favor of accurate geographic distances. 
All slopes and lifts are visualized as lines along their geographical location. 
Lifts are easily distinguishable due to their dashed line stroke and gray coloring, with an icon placed in the middle of the line indicating the lift type (\tBarLift,~\chairLift,~\gondolaLift,~\cableCarLift).
Traditional ski trail maps indicate slopes by colored lines in accordance with the local difficulty classification (see \autoref{fig:slopemap}), providing the user with a high-level overview of the resort. 
To better assess the gradient along the slope, we opted for a double-line visual approach consisting of an outer and inner line.
\jr{Similar to the slope marker poles which enclose the slopes on their sides, the outer line encodes the declared difficulty.}
The Austrian Standard~\cite{onorm} states that slopes where the steepness value does not exceed 25\% are labeled as easy \bluecircle, not exceeding 40\% as intermediate \redcircle, and slopes with even higher steepness values as advanced \blackcircle. 
Backcountry, or freeride slopes, which are typically ungroomed and tend to have a more extreme steepness profile, are often not indicated on the ski trail maps available from the resorts. 
We decided to include them in our design as they provide diverse alternatives for more experienced skiers.
Other mapping services~\cite{openskimap, fatmap} color these slopes in orange \orangecircle, a convention we adopted into our design.
Since the steepness was regarded as the most critical indicator regarding a slope's difficulty, we dedicated the inner line of our slope visualization to this feature. 
Every slope subsegment is colored according to the color scale in~\autoref{fig:colorlegend}, permitting a direct comparison between the difficulty classification (outer line) and the actual steepness (inner line). 
In addition, the introduction of the green color for negative steepness values allows easy identification of undesired uphill segments (\autoref{fig:slopec}).
In the case of ungroomed slopes, the line is displayed as pointed to mimic the appearance of a mogul slope (\autoref{fig:sloped}).
Due to the accurate representation of distances and angles in our map layout, the rough compass direction of every slope can be read directly from the map, and the contour lines provide information about altitude levels.

\begin{figure}[t]
    \centering
    \includegraphics[width=\linewidth]{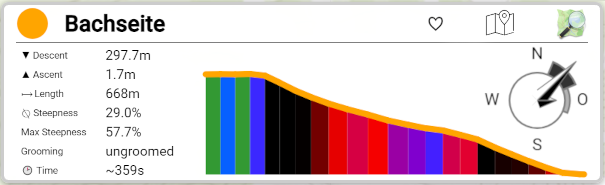}
    \caption{Tooltip displaying additional information about a slope. The altitude plot displays a flat, uphill section at the beginning of the slope, followed by a steeper section. The compass indicates the \visRev{north-east, east} direction of the slope. 
    }
    \label{fig:tooltip}
    \vspace{-.5em}    
\end{figure}

\subhead{Tooltip --} 
Additional information, such as the covered distance or the median travel time of each lift or slope can be obtained from a tooltip (see~\autoref{fig:tooltip}) accessible through a click on the respective line~\taskA. 
For lifts, details about the occupancy or seat heating are shown in a table-like enumeration.
Some Chairlifts provide seat heating or a bubble as protection from harsh weather conditions (e.g., snow or wind).
For slopes, the tooltip provides details about the ascent and descent, mean and max steepness as well as the type of grooming.
For a more detailed steepness investigation of the respective slope, an \textit{altitude plot} was integrated into the tooltip, a linechart where the x-axis encodes the length and the y-axis the altitude of the slope. 
For a visual comparison of the difficulty classification and the actual steepness, the line itself is colored according to the difficulty, while the area beneath follows our steepness color scale.
\jr{To leverage visual concepts familiar within the domain}, the compass direction distribution is depicted using a compass rose metaphor and grouped into the cardinal (N, E, S, W) and ordinal (NE, SE, SW, NW) directions. For each direction, a needle \jr{and an arc} are displayed on the rose, where the length of the needle \jr{and the grayscaled fill of the arc} encode the percentage of the slopes' subsegments pointing into the corresponding direction.



\begin{figure}[t!]
    \centering
    \includegraphics[width=\linewidth]{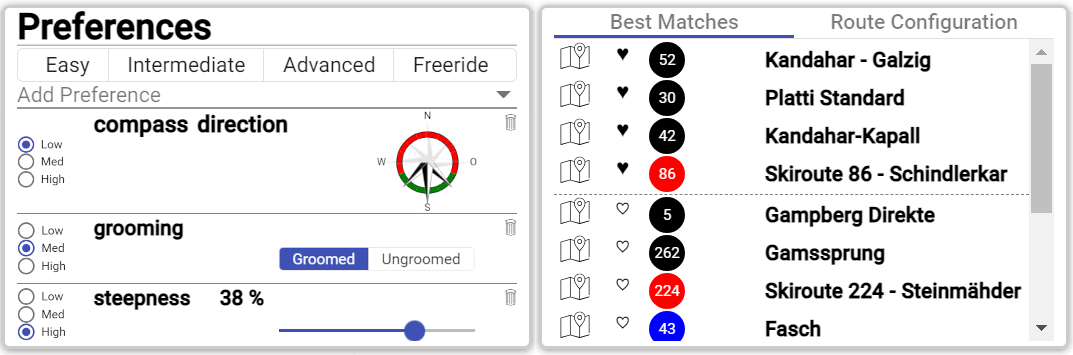}
    \caption{The user is interested in finding slopes that face \visRev{south}, are \visRev{groomed}, and exhibit a steepness of 38\%. The list of best matches returns a set of slopes that best match the given preferences, \visRev{where \protect\heartblack~indicates slopes that are part of the current recommended route}.}
    \label{fig:prefwidget}
    \vspace{-1.5em}
\end{figure}

\subhead{Integration of Preferences --}
An essential characteristic and major contribution of this work is the individualization of the workspace through the integration of user preferences~\taskB,~\taskD. A widget as seen in~\autoref{fig:prefwidget} allows users to input their weighted preferences towards the properties of \textit{steepness}, \textit{altitude}, \textit{compass direction}, \textit{grooming}, and \textit{crowdedness}. Default preference configurations for different difficulty categories (Easy, Intermediate, Advanced, or Freeride) are available and are selectable by a simple button press.
According to these preferences, slope recommendations are provided in a list in descending order of a computed score~\taskB.
The score ranges between 0 and 1 and is derived from the routing cost, which will be introduced in detail in the following paragraph. Since the total slope cost is computed as the sum of its subsegments' costs, the preference score $S_{pref}$ can be obtained using the following formula: $S_{pref} = (1 - (cost_{s}/K))$, where $K$ is the number of subsegments for a given slope.
Similar to the concept of Dunlop et al.~\cite{pistepreferences}, the slope visualization on the map is adapted whenever preference scores are updated, rendering preferable slopes and subsegments with greater width and higher opacity.

\begin{figure}[t!]
    \centering
    \includegraphics[width=\linewidth]{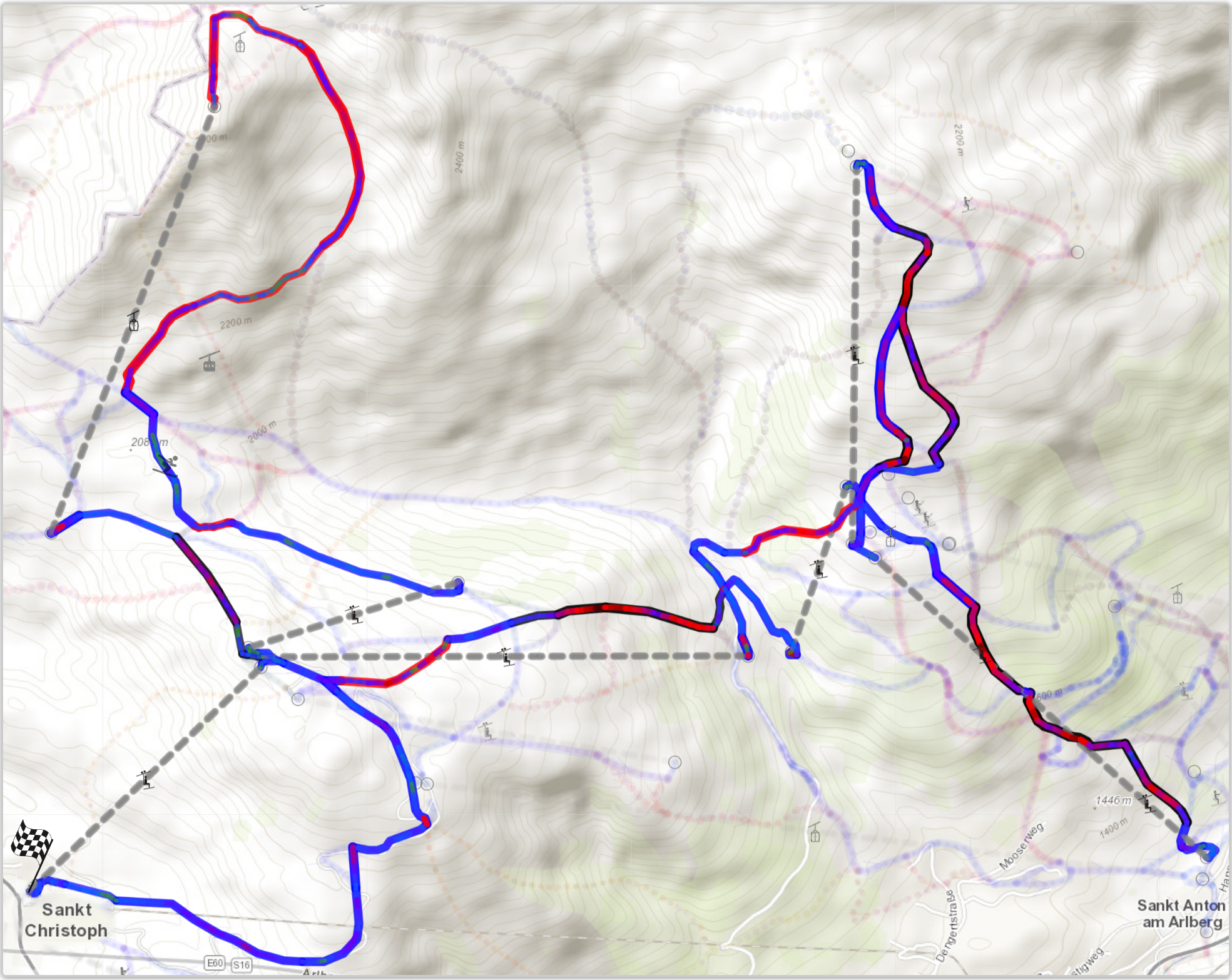}
    \caption{The recommended route can be highlighted on the map using opacity, providing a spatial overview of all slopes. However, the correct sequence in which the slopes have to be taken is unclear.
    }
    \label{fig:routeonmap}
    \vspace{-1.5em}
\end{figure}

\begin{figure*}[t!]
    \centering
    \includegraphics[width=\textwidth]{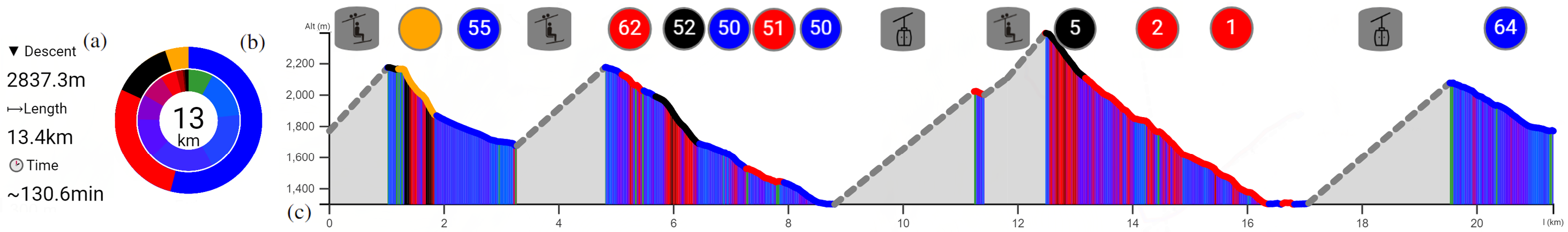}
    \caption{Route summary widget providing general route information (a). 
    The slope distribution donut chart (b) compares the slope difficulty (outer ring) and steepness (inner ring) distribution, revealing a dominance of \textit{easy} steepness segments \visRev{and only a small number of \textit{advanced} segments}.
    The altitude plot (c) sequentially visualizes the entire route and allows brushing to highlight the corresponding slope on the map. 
    }
    \label{fig:routesummary}
    \vspace{-1em}
\end{figure*}

\subhead{Routing --}
Given a routable network and two specific nodes, Dijkstra's algorithm~\cite{dijkstra} computes the shortest path connecting these nodes~\taskC. 
The shortness of the path is determined by costs which are assigned to each edge in the network according to specific properties. 
In accordance with~\cite{preferencebasedvehiclerouting}, we can leverage Dijkstra's algorithm to generate a route that is optimized according to the user preferences~\taskD~by adopting the cost for each slope with regard to the provided preferences of the user.
Given a set of preferences, each incorporating a weight and a value, the cost $cost_{s}$ for any slope is computed using the following formula:
\begin{equation}
    cost_{s} = \sum_{k=1}^{K} \frac{1}{P} \cdot (\sum_{p=1}^{P} w_{p} \cdot dist(v_{f}, v_{p}))\\
\end{equation}
    where 
    $K$ is the number of 30-meter subsegments belonging to the slope, 
    $P$ is the number of preferences with a weight > 0,    
    $w_{p}$ is the weight of a given preference, 
    $v_{p}$ is a given preference's value, 
    $v_{f}$ is a given subsegment's feature value, 
    and $dist(v_{f}, v_{p})$ is the distance of the actual feature value to the preference value. 
\jr{
 We employ the concept of attribute scoring functions (ASFs)~\cite{scoringfns:schmid:21} to compute the distance between the actual and the preference value.
ASFs generally transform the values of data attributes into numerical scores between 0 and 1 according to a mathematical function.
We choose a \textit{multi-point, continuous} transformation strategy for our numerical attributes, where we model a bell curve (normal distribution) with the mean as preference value $v_{p}$ and a standard deviation of 1.
For categorical attributes (\textit{compass direction} and \textit{grooming}), we apply a \textit{score assignment} approach, where desired values receive a score of $1$ and undesired values $0.1$ otherwise.}

Since slopes should generally be preferred over lifts, we assigned twice the maximum cost ($2 \cdot K$) to lift segments.
This ensures that bidirectional lifts that can be taken downhill are only taken when no other alternative is available (e.g., for linking lifts between resorts).

The recommended route can be visualized on the map (\autoref{fig:routeonmap}), where the slopes and lifts which are not part of the map are deemphasized with a lower opacity. 
In the case of a round trip, or if the same slope is taken multiple times, the traceability of the route can be compromised. 
To improve the traceability regarding temporal ordering, we implemented a \textit{route summary} widget where the entire route is visualized in an altitude plot (\autoref{fig:routesummary}). 
Hovering along the altitude plot with the mouse places a skier icon~\skierIcon~on the corresponding slope segment on the map, allowing the user to easily follow the itinerary of the route. 
Slope sign symbols similar to the actual signs along the slope are added above the altitude plot to support the navigation when following the slope signs.
A double donut chart provides distribution information about both the difficulty classification (outer ring) and the steepness of the segments (inner ring), allowing to visually compare these attributes.
General route details, namely the vertical descent, the total length, and the estimated time, are shown on the left-hand side.



\subhead{Workflows --}
In consultation with ski guides and enthusiasts, it became apparent that, depending on the level of familiarity, different strategies exist when manually planning routes in ski resorts. 
If skiers already have prior knowledge of the resort, they might have already determined a particular set of favorite slopes. In that case, they would like to receive precise recommendations on how to visit all of these slopes.
Otherwise, if skiers are visiting the resort for the first time, they might be overwhelmed at first and would like to effortlessly receive recommendations based on their preferences. 
\visRev{Similar to the \textit{end-to-end} and \textit{step-by-step} strategies introduced in~\cite{tourplanning:deng:2022},} we developed two workflows for route recommendation to meet both of these approaches:

\textbf{Automated workflow}: For unfamiliar resorts, users input their preferences, a start and end location, and the duration of the desired route. 
According to their preferences, matching slopes in proximity will be selected and sequenced using a traveling salesman approach.
The route between these slopes will then be calculated, resulting in the final route. 
The users receive the route and the slopes selected as their favorites, allowing them to readjust the initially recommended route.

\textbf{Semi-automated workflow}: If the users are already somewhat familiar with the resort, they select matching slopes themselves by clicking the \heartwhite~symbol either from the tooltip or the list of best matches (see~\autoref{fig:prefwidget}), effectively performing the first part of the aforementioned workflow manually. 
The list of best matches serves as a pool for recommended slopes, albeit users can also take advantage of their resort knowledge and select their favorite slopes from the map directly.







\section{Evaluation}
\visRev{To evaluate the applicability of \skivis, we provide an exemplary usage scenario leveraging both workflows, as well as a detailed user study.}

\subsection{Usage Scenario}

\visRev{Jamie is an experienced skier who annually partakes in skiing trips. This year, she visits \textit{St. Christoph} in the \textit{Ski Arlberg} region for the first time and uses~\skivis~to guide her through the resort. 
As a first step, she selects the \textit{Advanced} preference configuration to explore the resort~\taskA.
From the list of best matching slopes, she can identify several slopes fitting her criteria~\taskB.
Being unfamiliar with the resort, she initially opts for the \textit{automated workflow} and desires a route recommendation of two hours duration for a morning run before returning to the hotel for lunch~\taskD. 
Driven by her domain knowledge, she sets her preferences on compass direction to southern-facing slopes, as she prefers to ski on sunny slopes.
\autoref{fig:prefwidget} shows the provided preferences and the slopes considered as favorites by the routing algorithm (indicated by \heartblack).
The recommended route is seen in~\autoref{fig:routeonmap}.
%
For the afternoon, she decides on the \textit{semi-automated workflow} to generate a trip of around two hours duration~\taskC. 
She enjoyed the \textit{52 Kandahar-Galzig} slope from her morning run and marks it as one of her favorites as she wants to ski it again. 
She also selects the \textit{5 - Gampberg Direkte} slope, a slope from the list of best matching slopes that did not make it into the \textit{automated} route suggestion she followed in the morning. 
She is further looking for an ungroomed, north-facing slope, knowing the snow conditions are better preserved with lesser sun exposition. 
By adjusting her preferences accordingly, she identifies a number of possible candidates, and finally decides on the \textit{Bachseite} slope seen in~\autoref{fig:tooltip}, as it does not include any uphill segments and adheres to the time restrictions. 
With an estimated duration of 130 minutes, she finalizes the planning with the route visible in~\autoref{fig:routesummary}.
}

\subsection{User Study}

\visRev{For a quantitative, task-based evaluation of~\skivis,} we conducted online pair analytics sessions \cite{pairanalytics} with six experts from the ski domain.

\subsubsection{Study Design}
Based on the task requirements established in~\autoref{section:requirements} and workflows introduced in~\autoref{section:application}, the participants had to carry out three analysis challenges explained in the study methodology.
The feedback from the participants regarding their experience with the interface serves as the foundation for identifying the benefits and drawbacks of our approach. 

\subhead{Participants --}
We defined the target audience of a visual exploration and route planning interface in~\autoref{section:requirements}. Hence, we selected our study participants accordingly. 
In total, six participants with at least intermediate (>ten years of experience) skiing expertise and varying knowledge about the resort Ski Arlberg partook in the study.
Three participants were certified ski guides in the resort and are further referenced as \guideA~--~\guideC. 
Their extensive knowledge of the slopes is valuable when evaluating the quality of the recommended routes.
The remaining participants were intermediate or advanced skiers with little or no knowledge about the resort and are further referenced as \skierA~--~\skierC.

\subhead{Methodology and Tasks --}
The study procedure comprised three sections through which the participants were guided by a \textbf{v}isual \textbf{a}nalytics \textbf{e}xpert (VAE) of our team. 
The duration of each trial was scheduled for 60 minutes, with an average duration of 68 minutes for all study trials.

The first part of the study marked a general introduction to the domain and the tasks at hand. 
Each participant filled out a demographic questionnaire recording their age, gender, profession, skiing proficiency, the average number of skiing days per year, knowledge of the Ski Arlberg resort, and whether they possessed a skiing instructor license. 
Several further questions were asked to assess their knowledge on resources that facilitate visual exploration or route planning in ski resorts and their frequency of use of these resources.
We further inquired about their main criteria when planning routes, as well as their general expectations regarding an interactive workspace for visual slope exploration and route planning in ski resorts. 

In the second part, the participants were given a detailed explanation of all components of the study prototype by the VAE. 
Afterward, the participants were given the link to access the web-based prototype to acquaint themselves with the application and interaction methods. 
Their first task was to explore the resort independently~\taskA~and investigate the locations and characteristics of the slopes that were recommended to them according to their applied preferences~\taskB.
After the initial investigation, their next task was to generate a route (\textasciitilde five hours duration) incorporating these preferred slopes from a start and end location of choice~\taskC, following the \textit{semi-automated workflow}.
As their last task, the participants were asked to plan a route without favoring any specific pistes (based on the \textit{automated workflow}), simply by providing the same start and end locations, the duration of the route (again \textasciitilde five hours), and their applied preferences~\taskD.

The third and final part of the study consisted of an interview where the participants were asked about their experiences, insights, and improvement suggestions and to what extent \skivis~met their initial expectations.
We concluded the study by assessing the participants' conceptions regarding the mental, physical, and temporal demand through the NASA Task Load Index (TLX)~\cite{nasatlx}. 

\subsubsection{Evaluation Results}
In the following section, we summarize the feedback received from the domain experts during the study. 
We structured the section concerning the participants' initial expectations, their opinions on the exploration and route generation tasks, and further possible use cases for~\skivis~that emerged during the final discussion phase. 

\subhead{Expectations --}
The responses toward the initial expectations varied among the participants. 
~\guideC~envisioned a Google Maps for ski resorts, where he would receive detailed routing instructions for navigation between locations within a resort. 
Aside from slope routing,~\guideA~and~\guideB~expected ski huts and restaurant information as part of the analysis process.
\skierB~and~\skierC~anticipated information regarding traffic on the slopes as well as waiting times at lifts. 
Furthermore,~\skierA~expected a social network functionality where the experiences of other skiers regarding lifts and slopes are shared.
Regarding preferences, \guideA~and~\skierA~stated that they expect the route planning to incorporate constraints regarding the difficulty classification (e.g., to only ski on blue slopes).



\subhead{Exploration and Workflow Interaction --}
A finding that multiple participants observed in the exploration task was that many of the slopes in the ski resort exhibit segments where the steepness deviates from the difficulty classification.
The possibility of visually exposing uphill sections along the slopes through the green color (see~\autoref{fig:slopec}) was also valued by multiple participants. 
\guideA~and~\skierC~were interested in detecting north-facing slopes (since better snow conditions are better preserved at later times of the day), and were able to identify regions where these slopes were predominantly present. 

All study participants mentioned that they prefer the \textit{automated workflow} when planning routes for a completely unknown resort. With increasing familiarity, the benefit of selecting individual, known slopes and planning the route around these favorites outweighs the shorter execution time, resulting in a preference for \textit{semi-automated workflow}. 
The functionality to receive different route recommendations depending on the preference settings was appreciated by \guideB~and \guideC, allowing them to tailor the route planning to their customers' needs.
\skierA~labeled the route summary (see~\autoref{fig:routesummary}) as the \emph{table of contents} of the ski day and emphasized the benefit of linking the visualization to the map on hover.
Regarding the recommended routes,\jr{~\guideA~and~\guideC~concurred with the presented suggestions, agreeing that they are appropriate choices for their selected preferences.
}
\guideB~expressed skepticism about freeride~\orangecircle~recommendations. Particular caution should be taken on these slopes, as external factors such as avalanche dangers, wind, or visibility conditions can complicate the downhill ride or even rule it out completely. Blindly following the system's recommendations could lead inexperienced skiers into hazardous terrain.

\guideC~suggested that the routing capabilities could be extended to integrate bus routes. 
Since Ski Arlberg is the largest resort in Austria, it is comprised of several towns which are also connected by buses.
Since the expectation of \guideA~and~\guideB~was to receive information about restaurants or huts, they suggested integrating such information into the prototype. 
Especially when planning routes for an entire day, the possibility of including a stop for lunch was seen as an enrichment to the route planning process.
Although the routing capabilities were seen as a helpful feature to plan a ski day, \guideB~and \guideC~critizised that finding the correct slope when actually skiing is a challenge that is not addressed by the application.

\begin{figure}[t!]
    \centering
    \includegraphics[width=\linewidth]{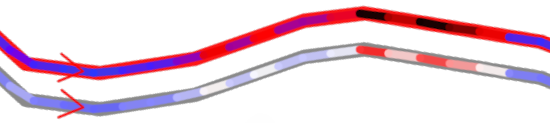}
    \caption{Two visualization variants for a slope classified as intermediate, containing segments that deviate significantly in steepness. 
    The upper slope is colored accordingly to the design introduced in~\autoref{section:application}. The lower slope emphasizes steepness deviations from the  difficulty classification (25\%--40\%) for intermediate.
    }
    \label{fig:discrepancy}
\end{figure}

\subhead{Derived Use Cases --}
During the study, multiple participants mentioned further application areas for \skivis, which we outline in the following section. \jr{Since these feature requests surfaced during the study, they were implemented afterward and thus not evaluated by our experts.}

According to the finding that the steepness of the slope often deviates from the difficulty classification, the feature to assess the entire resort regarding this discrepancy was elicited as a further use-case scenario by multiple study participants. 
To visually support this assessment, we designed an alternative color scale that encodes the \emph{difference} between the actual steepness of the segment and the value ranges used in the Austrian standard for difficulty classification~\cite{onorm}. We propose a diverging scale from blue over white to red, representing segments that are less steep than declared, as steep as declared, and steeper than declared, respectively. \autoref{fig:discrepancy} compares this novel coloring scheme to the one established in~\autoref{section:application}. 
As blue and red are established colors for indicating easy and intermediate slopes, respectively, we are aware of the confusion this coloring introduces. 
However, in this setting, we aim to showcase segments that deviate from their difficulty classification.
The novel coloring scheme leverages the association of red with danger~\cite{reddanger:pravo:14}, accentuating slope segments that are steeper than they are actually indicated on available trail maps.

\begin{figure}[t!]
     \centering
     \begin{subfigure}[b]{0.24\linewidth}
         \centering
         \includegraphics[width=\textwidth]{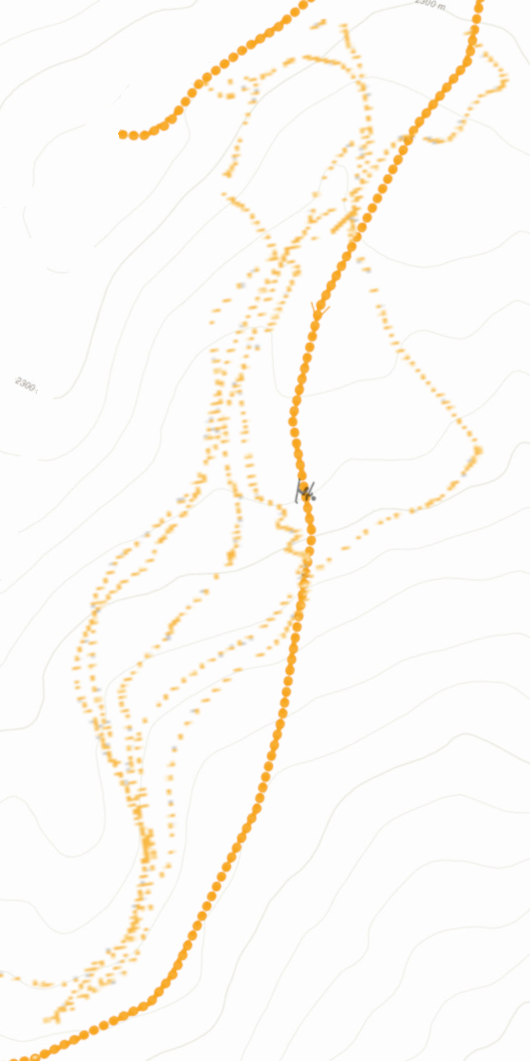}
         \caption{Freeride trajectories.} 
         \label{fig:traj_free}
     \end{subfigure}
     \hfill
      \begin{subfigure}[b]{0.24\linewidth}
         \centering
         \includegraphics[width=\textwidth]{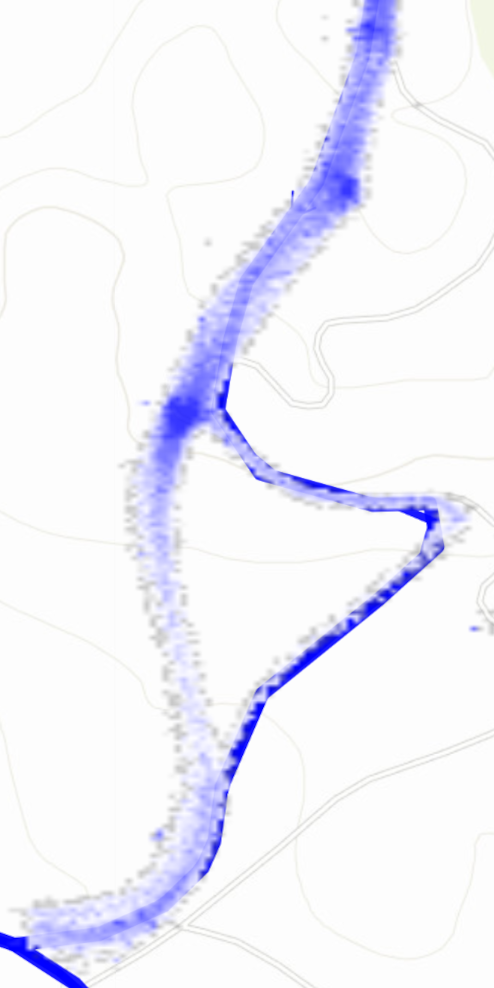}
         \caption{Stop points along a slope.}
         \label{fig:traj_stop}
     \end{subfigure}
    \hfill
     \begin{subfigure}[b]{0.24\linewidth}
         \centering
         \includegraphics[width=\textwidth]{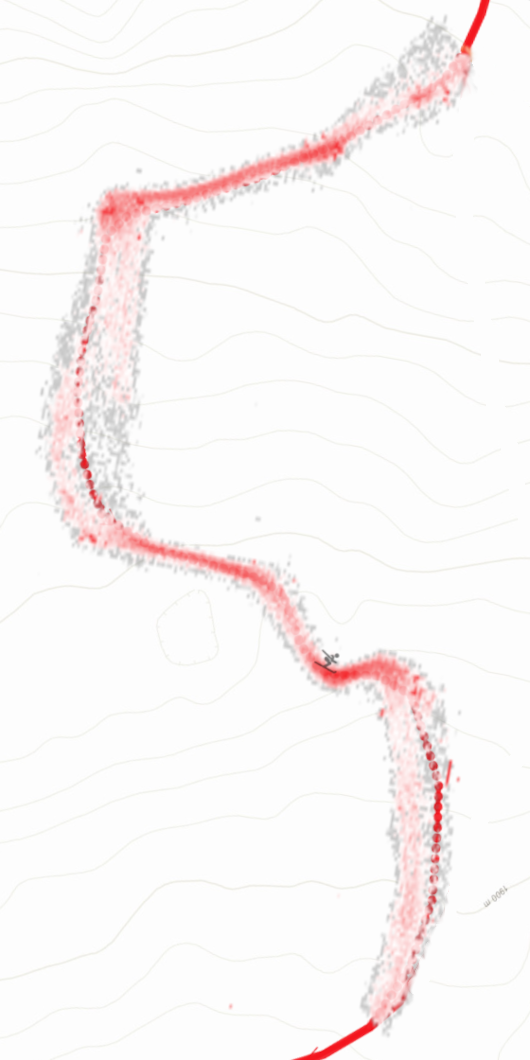}
         \caption{Estimation of slope width.}
         \label{fig:traj_width}
     \end{subfigure}
     \hfill
     \begin{subfigure}[b]{0.24\linewidth}
         \centering
         \includegraphics[width=\textwidth]{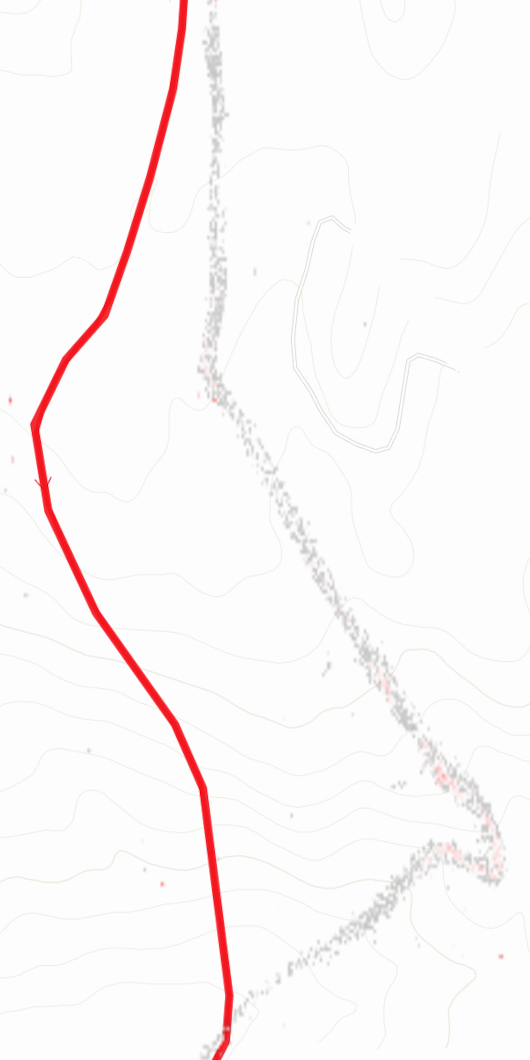}
         \caption{Mismatch of OSM and trajectories.}
         \label{fig:traj_wrong}
     \end{subfigure}
     \hfill
        \caption{Visualization of actual trajectories along the slopes, revealing several findings.}
        \label{fig:heatmap}
        \vspace{-1.5em}
\end{figure}

Another requested feature that surfaced during the discussion phase was the analysis of actual skier behavior, more precisely \jr{specific habits such as stopping points} or out-of-bounds skiing. 
\jr{
It is not uncommon to take a short break along parts of a slope. 
Especially when skiing in larger groups, such breaks occur more often when faster group members have to wait for slower members to catch up.
Care must be taken to remain visible and not hinder any traffic when selecting a location to stop (i.e., on the side of the slope on a flat segment).
Otherwise, other skiers might overlook a waiting party, resulting in a crash and potential injuries.
Identifying and analyzing common stop points can be beneficial for resort planners, as they can shape the slopes accordingly to minimize injury risk.
}
Furthermore, more experienced skiers often seek possibilities to leave the marked slopes and enjoy untouched, deep powder snow.
\jr{
While ungroomed freeride slopes typically offer powder conditions after a night of fresh snowfall, further options open up when leaving the marked slopes and venturing out-of-bounds.
}
Extensive knowledge of the ski resort is required to locate such unmarked trails, assess their safety, \jr{and ensure a return to the marked slopes without tedious uphill walking is possible}.

\jr{
Such investigations can be conducted by visualizing the recorded GPS trajectories, as selected examples presented in~\autoref{fig:heatmap} showcase. 
As displaying individual raw trajectories would not scale, we calculated a spatial kernel density estimation of the trajectory points and generated a raster image representing each slope's traffic intensity as a heatmap.
Recommendations for skiing out-of-bounds can be taken from single trajectory paths, as observed in~\autoref{fig:traj_free}.
Stop points can be identified as hotspots, visible in~\autoref{fig:traj_stop}.
Aside from the above-elicited use-cases, this technique allows the examination of other interesting causes.
As~\autoref{fig:traj_width} demonstrates, the visualization enables an approximation of the width of individual slopes, which is a property not included in the OSM data.
The uncertain data quality of the volunteered data from OSM can visually be assessed with the trajectory heatmap. In~\autoref{fig:traj_stop}, a shortcut taken by many skiers can be identified that is not recorded. \autoref{fig:traj_wrong} reveals a vastly different path for a certain slope.
}




\section{Discussion}
\label{section:discussion}
The study revealed several benefits and drawbacks of our implemented approach. 
This section discusses the major controversial issues, outlines limitations, and provides an outlook on further research opportunities.
\jr{
\skivis~provides a novel perspective to view and interact with ski resort data.
Many observations made by the participants, such as detecting uphill segments or north-facing slopes, are not supported by the current state of the art.
The route planning capabilities can aid skiing enthusiasts in centering their ski day around their preferred slopes, even when navigating unfamiliar territory. 
A concern raised by several study participants was that the dangers of avalanches were neglected in~\skivis, especially regarding off-piste ventures.
Thus, we extended our prototype to display a disclaimer when a freeride slope is contained in the route recommendation to address this.}

\subhead{Color Coding --}
The color scale used to encode steepness values (\autoref{fig:colorlegend}) incorporates both red and green color hues. 
Research suggests that red-green color blindness is the most common vision deficiency~\cite{colorblindness:Birch:12}. 
Furthermore, some countries (e.g., France, Norway, or the United States) employ a different color-coding system on their slopes, where green slopes are sometimes classified as \textit{very easy}. 
\jr{As we limit our case study to an Austrian ski resort, we aligned the colors to the prevailing system.
However, the coloring can, in theory, be adapted to follow any local scheme. } 
\guideA--\guideC~did not voice any concerns when confronted with this discrepancy and further expressed that the color scale helped them to identify undesired flat segments, albeit~\guideC~and \skierC~noted that the visual design appeared almost too colorful.
\jr{We implemented the option to reduce the design presented in~\autoref{fig:designspace} to only the outer line, to reduce the cognitive load of users who prefer a more simple design.}

\subhead{Weather Information --}
Multiple study participants addressed the absence of information about weather conditions. 
Providing accurate weather predictions is a vast research topic on its own~\cite{weatherpredictionsurvey}, and the reliability of such information decreases rapidly the further one attempts to forecast into the future, especially in mountain regions. 
While a \textit{context-aware} routing algorithm as proposed by~\cite{preferencebasedvehiclerouting} is capable of incorporating such settings, weather information is given for the entire resort and thus cannot be subdivided on individual slopes to serve as decision-making criteria.
However, by using domain knowledge, accurate daily weather information from external sources can be utilized to derive user preferences. In case of strong winds or heavy snowfall, high altitudes should be avoided, which can be achieved by setting the altitude preference to a lower value. 
Similarly, slopes that receive high sun exposure tend to exhibit difficult snow conditions later in the afternoon and are evaded by preferring north-facing slopes.

\subhead{Map Design --}
\jr{
In contrast to our 2D topographic map design, panorama maps are the dominant style to render ski trail maps~\cite{skimaps}. 
Aside from facilitating navigation and orientation, visual aesthetics and marketing reasons heavily influence their design process. 
While Field~\cite{skimetromap} proposed an alternative representation based on a metro map layout, the evaluation concluded that the panorama style is too established and familiar for alternative concepts to succeed.
However, a study by Balzarini et al.~\cite{deformedmap} revealed that skiers feel uncertain and face problems in understanding when performing tasks such as locating, wayfinding, or decision-making using a panorama map.
The deformation of the terrain makes it impossible to display distances and compass directions at any given location accurately.
These drawbacks cannot be reconciled with our requirements, specifically the exploration of a ski resort according to geographic features~\taskA.
Since metro maps apply abstractions to simplify the underlying geography~\cite{metromapsurvey:wu:20}, they are also unsuitable for our purposes.
While 3D maps are advantageous for understanding elevation differences~\cite{2d3dtrailheadmap:schobesberger:08}, their shortcomings include perspective distortion and occlusion issues, depending on the viewpoint~\cite{cartography:buckley:04}.
A study evaluating a 3D environment for hiking trails suggests that 2D map layouts are preferred for route detection and planning tasks and that participants had trouble assessing distances and identifying the steepest section in a 3D layout~\cite{virtualhikeplan:bleisch:08}.
These findings motivated us to deploy a 2D map layout, although the option to provide both options (as done by~\cite{osmskirouting}) would be a holistic compromise.
During the study interviews,~\guideC~and~\skierC~appreciated the contour lines as a help to assess the terrain steepness. 
\guideB~stated that a 2D layout is sufficient for piste navigation, although 3D terrain visualizations would be helpful for off-piste ventures.
Aside from the contour line approach, we provide other tiling layers, such as relief shading, that can be interactively selected. 
}

\subhead{Data Quality --}
\jr{
The quality of VGI data is often evaluated by comparing it against a verified dataset released from official sources~\cite{vgiquality:fonte:17}, which is problematic since no or little official information is publicly available. 
The ISO 19157 standard~\cite{isogeoquality:iso:13} defines data quality for geographic information, on which we base this assessment on. }

\jr{\textit{Thematic accuracy} evaluates the correctness of the OSM tags, namely the \textit{difficulty} and the \textit{grooming} of each slope.
The difficulty can be assessed by comparing it against the official slope map~\cite{arlbergmap}, whereas clear indications about the grooming status are not given. 
OSM tags comprise easy, intermediate, advanced, and freeride values, whereas Ski Arlberg classifies their slopes into easy~\bluecircle, medium~\redcircle, challenging~\blackcircle~\blackdiamond, and skiroute~\reddiamond, with the diamond shape presumably indicating partly ungroomed slopes.
We noticed that these slopes were tagged by their respective color (\reddiamond~as intermediate and~\blackdiamond~as advanced) and marked as ungroomed in OSM, with two exceptions (3.5\%) which were tagged as freeride slopes.
Aside from these irregularities, no slope marked as freeride in the OSM data is visible on the official slope map.
}

\jr{The \textit{positional accuracy} of both OSM and extracted trajectories can be assessed visually in our prototype by comparing them against each other, as~\autoref{fig:heatmap} illustrates.
However, errors from GPS transmitters, which can occur in mountain regions, cannot be ruled out completely.
}

\jr{In terms of \textit{completeness}, details about the total length of slopes by difficulty are available~\cite{bergfex}, which can be compared against our slope network.
Individual missing or excess slopes cannot be detected, thus only serving as a broad indicator.
\autoref{tab:bergfexvsskivis} indicates slight discrepancies between the numbers, which are to be expected given the above-mentioned disparities.
The recorded GPS trajectories undeniably do not fully cover the entire extent of ski resort visitors. 
Yet, we argue that our sample is significantly large enough to draw estimations about \textit{crowdedness} and \textit{travel durations} values. 
Nonetheless, related research indicates that socioeconomic biases can be contained~\cite{stravabias:zander:2016}, and we assume more experienced skiers record their rides more frequently, although we are unable to prove this scientifically.}
\begin{table}[]
    \small
    \centering
    \begin{tabular}{|c||c|c|c|c|}
        \hline
        Source & \bluecircle~easy & \redcircle~intermediate & \blackcircle~advanced & total\\ \hline \hline
         Bergfex & 130km & 122km & 51km & 302km \\
         \hline
         \skivis & 134km & 114km & 46km & 294km \\
         \hline
    \end{tabular}
    \caption{Total slope length by difficulty.}
    \label{tab:bergfexvsskivis}

    \vspace{-2.5em}
\end{table}

\subhead{Transferability --}
\jr{
To showcase the potential of our approach, we focused on the region of Ski Arlberg, being the largest ski resort in Austria.
However, our methods are in principle resort agnostic and applicable if the requirements introduced in the following paragraph are met.
A \textit{graph data structure} consisting of the resorts' slopes and lifts is a necessity to facilitate a general routing algorithm.
Spatial information of every edge of said graph is required in combination with a \textit{DEM} to derive geographic properties such as the altitude, steepness, and compass direction.
Additional properties (such as difficulty or grooming) aside from the spatial component refine the preference-based visualization and routing capabilities. 
We used OSM to obtain the resorts' graph network structure, which provides volunteered geographical details for many other ski resorts worldwide.
However, other officially verified sources (if available) can be used as well to improve the data quality.
Aside from Copernicus~\cite{copernicus}, other DEMs can be employed to acquire elevation details.
At last, \textit{GPS trajectories} from the resort are required to estimate popularity and travel durations and perform skier behavior explorations.
While the proposed methods are not limited to a certain resort size, larger resorts benefit more from the methods presented since a route planning feature is not necessarily needed for smaller resorts with only three lifts.
With the mentioned data requirements available, our concepts can also be transferred to related domains such as hiking or mountain biking.
}


%


\subsection{Limitations}
While the study manifested the capabilities of \skivis~in regard to the elicited tasks, some limitations became apparent.
For example, the visualization and analysis methods deployed in the application heavily emphasize the steepness property, and lift preferences (e.g., to exclude t-bar lifts~\tBarLift) are not integrated into the routing algorithm.

\begin{wrapfigure}[4]{r}{0.35\linewidth}
    \begin{center}
    \vspace{-.6cm}
    \includegraphics[width=\linewidth]{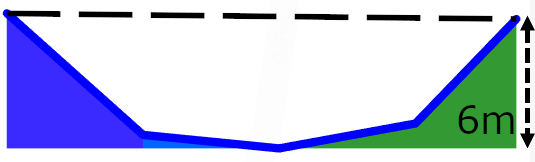}
    \end{center}
\end{wrapfigure}
DEMs provide altitude information based on the underlying geographic topology.
In exceptional cases (such as bridges or tunnels), the slopes do not strictly follow this topology, resulting in incorrect steepness calculations. 
\jr{Since the recorded GPS trajectory data also includes an altitude property, a comparison between the DEM and the altitude information of the matched trajectories could possibly expose such discrepant cases and could prove to be a viable strategy to mitigate this limitation.}

\begin{wrapfigure}[4]{r}{0.35\linewidth}
    \begin{center}
    \vspace{-.6cm}
    \includegraphics[width=\linewidth]{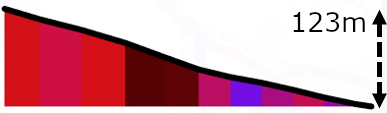}
    \end{center}
\end{wrapfigure}
The y-axis of the \textit{altitude plot} is always scaled according to the minimum and maximum altitude, meaning the visual slope of the individual segments is dependent on the total vertical descent of the respective slope.
Thus, the steepness is perceived as more severe for slopes with little vertical descent. 
This limitation becomes apparent when comparing two plots where the total vertical descent largely deviates.
However, the color-coding of the segments below the line still indicates the actual steepness. 


Although we dedicated a stage in the preprocessing to improve the connectedness of the network and connect spatially close, unconnected nodes, some slopes still remain unconnected and constitute a total of 15 dead ends in the network topology.
These slopes cannot be integrated into the routing, as a return path to other slopes is not available.
Most of these dead ends are located along roads, and including bus routes (as suggested by~\guideC) could further reduce the number of such impasses.

\jr{
\skivis~provides a time approximation for each recommended skiing route.
The current solution calculates the median travel time of recorded trajectories on a given slope and assumes the sum of all slopes in the route to be the required duration.
This is a first approximation but arguably imprecise and generalized since skiers with varying levels of expertise ski at different speeds. 
While an exact number is very hard to obtain, sophisticated machine learning techniques could be deployed to better and individually estimate a skier's speed and consequently improve the duration estimation of the routing algorithm.
}

\guideB~pointed out that the routing recommendations help to find preferred slopes and plan out an entire day of skiing, yet are not beneficial for actual orientation on the mountain. 
A mobile-optimized version would allow users to use the prototype on their smartphone but not facilitate the transfer of the visualized route to the real world. 
The development of embedded systems is beyond the scope of this research.
However, novel AR techniques~\cite{ridersguide:schmid:11 ,skiar:fedosov:16, vrmaptransform:ghaemi:22} or wearable technology could be the solution to guide the skier along the recommended route and indicate the correct slope to take when approaching a junction.

An entire day of skiing usually includes at least one break. Hence,~\guideA,~\guideB, and~\skierB~requested integrating ski huts and restaurants into the routing algorithm. 
While the interactive procedure of planning a route by selecting favorite slopes was appreciated by the study participants,~\skierA~desired the possibility to blacklist certain unwanted slopes. Providing alternative bypasses on demand could further enhance the routing capabilities.
Additional features, such as the \textit{curvature} or \textit{width} of the slope, could refine the capturing of preferences and result in a better-personalized route.

\subsection{Future Work and Research Opportunities}
As visual analytics methods in the ski domain are underexplored, we propose this work as a starting point, urging further research. 


\subhead{Context-aware Routing --}
\jr{As desired by multiple study participants, integrating dynamic properties such as weather or snow conditions would further benefit the routing capabilities. 
The high uncertainty and volatility of such data pose challenges and can question reliability.
Nonetheless, promising insights could be obtained, such as finding the weather-optimal time for a lunch break or optimizing a route to ski on the best snow conditions for the entire day.}


\subhead{Temporal Resort Dynamics --}
Our approach to estimate the crowdedness of slopes only permits a coarse insight into the travel activities within a ski resort. 
With temporally more fine-grained trajectory data, a sophisticated model could identify cyclic patterns about slope crowdedness and lift congestion (e.g., certain slopes exhibit high traffic only during morning hours).
Broader temporal trends such as (un)popular weekends could be explored to determine an optimal time interval with empty lifts and slopes.
Slope conditions also undergo substantial changes throughout the day and the entire season.
eHyd~\cite{ehyd:Austria:22} provides details on snow heights and new snow for various locations in Austria and could serve as a starting point to investigate temporal changes.

\subhead{Collaborative Route Planning --}
\jr{
Recent research~\cite{skigroups:delibasic:22} validates the claim that skiing in a group is more popular than skiing alone, \visRev{thus further complicating the route planning procedure as preferences from different users can contradict each other.}
This constitutes the need for a collaborative route planning environment in which multiple users can contribute interactively to the exploration and route generation process.
To accommodate the diversity in preferences of the users, \visRev{a comparison of different recommendations would benefit the decision-making process.
Guidelines for the task of visual comparison~\cite{viscompare:gleicher:18, complayouts:lyi:21} exist and recommend the use of superposition~\cite{businesscomp:wang:18} when the spatial context is considered as significant.
}
A common meeting point, such as a specific restaurant at lunch, could also be a sustainable compromise. 
}

\subhead{Community-Based Annotation Methods --}
\jr{
A limitation of all ski trail maps so far is that they cannot be enhanced with content from the users themselves.
Fedsov et al.~\cite{skiar:fedosov:16} already experimented with sharing information on HMDs, revealing that the participants appreciated the functionality to mark hazards or points of interest. 
A larger-scale social network (as envisioned by \skierA) could establish a geographic information-sharing platform in ski resorts.
The potential use cases could range from general suggestions, such as slope quality, vantage points, or restaurant recommendations, to more temporal dependent information, such as snow conditions, accidents, slope closings, or long lift queues.
}

\subhead{Resort comparison --}
\jr{
Our work is limited to the analysis of the Ski Arlberg resort. However, our approach is transferable to any other ski resort.
Austria alone is home to 253 resorts with five or more lifts, 15 of which generate more than 1 million skier visits per season~\cite{skitourismreport:vanat:21}, complicating the search of finding the most suitable resort. 
Providing a detailed comparison between resorts in regard to decisive attributes such as price, travel time, resort size, altitude, and family-friendliness can support ski enthusiasts in choosing a resort.
The difficulty classification of slopes varies among countries and resorts, and a comparison and assessment of which factors contribute to which extent could expose discrepancies and pave the way for uniform guidelines.
}







\section{Conclusion}

In this paper, we introduced~\skivis, an interactive workspace for visual exploration and route planning in ski resorts.
We identified requirements and task characteristics for such an application, 
and conducted pair analytics studies with six experts from the ski domain to evaluate the applicability of our solution.
The qualitative study results demonstrate that skiers can benefit from interactive visualizations.
The visual slope exploration allowed the experts to detect patterns such as uphill segments, 
and the preference-based routing recommendations were highly appreciated when investigating unknown ski resorts.
While we discuss the benefits and limitations of our approach, we furthermore present research opportunities and encourage fellow researchers to engage in the rather unexplored field of visual analytics for ski resorts.  

\clearpage

\acknowledgments{
This work was funded by Deutsche Forschungsgemeinschaft (DFG) under grant 455910360 (SPP-1999) and the priority programme\linebreak "Volunteered Geographic Information: Interpretation, Visualisation and Social Computing” (VGIscience, priority programme 1894).
}

\bibliographystyle{abbrv-doi-hyperref}

\bibliography{references/refs}

\begin{thebibliography}{10}

\bibitem{dynamicpricing:alnes:21}
P.~K. Alnes, I.~Malasevska, Ørjan Mydland, and E.~Haugom.
\newblock {Price differentiation in the alpine skiing industry—The challenges
  of demand shifting and capacity constraints under pandemics}.
\newblock {\em J. Outdoor Recreat. Tourism}, 35:100409, 09 2021.
  \href{https://doi.org/10.1016/j.jort.2021.100409}
{doi: {{%
10\hspace{.1pt}\discretionary{.}{%
}{.}\hspace{.4pt}1016\discretionary{/}{%
}{/}j\hspace{.1pt}\discretionary{.}{%
}{.}\hspace{.4pt}jort\hspace{.1pt}\discretionary{.}{%
}{.}\hspace{.4pt}2021\hspace{.1pt}\discretionary{.}{%
}{.}\hspace{.4pt}100409}}}


\bibitem{pairanalytics}
R.~Arias-Hernandez, L.~T. Kaastra, T.~M. Green, and B.~Fisher.
\newblock {Pair Analytics: Capturing Reasoning Processes in Collaborative
  Visual Analytics}.
\newblock In {\em Int. Conf. Syst. Sci.}, pp. 1--10, 2011.
  \href{https://doi.org/10.1109/HICSS.2011.339}
{doi: {{%
10\hspace{.1pt}\discretionary{.}{%
}{.}\hspace{.4pt}1109\discretionary{/}{%
}{/}HICSS\hspace{.1pt}\discretionary{.}{%
}{.}\hspace{.4pt}2011\hspace{.1pt}\discretionary{.}{%
}{.}\hspace{.4pt}339}}}


\bibitem{onorm}
{Austrian Standards Int.}
\newblock {ÖNORM S 4611 - Signs for use in organised skiing areas -
  Requirements, design and classification}, 2016.

\bibitem{deformedmap}
R.~{Balzarini}, A.~{Dalmasso}, and M.~{Murat}.
\newblock {A Study on Mental Representations for Realistic Visualization the
  Particular Case of Ski Trail Mapping}.
\newblock {\em ISPRS Int. Arch. Photogramm., Remote Sens. Spatial Inf. Sci.},
  XL3:495--502, 2015.
  \href{https://doi.org/10.5194/isprsarchives-XL-3-W3-495-2015}
{doi: {{%
10\hspace{.1pt}\discretionary{.}{%
}{.}\hspace{.4pt}5194\discretionary{/}{%
}{/}isprsarchives\discretionary{%
}{-}{-}XL\discretionary{%
}{-}{-}3\discretionary{%
}{-}{-}W3\discretionary{%
}{-}{-}495\discretionary{%
}{-}{-}2015}}}


\bibitem{bergfex}
{Bergfex GmbH}.
\newblock {Bergfex}.
\newblock \url{https://www.bergfex.at/}, 2022.
\newblock Accessed 27.06.2023.

\bibitem{colorblindness:Birch:12}
J.~Birch.
\newblock Worldwide prevalence of red-green color deficiency.
\newblock {\em J. Opt. Soc. Am. A}, 29(3):313--320, 03 2012.
  \href{https://doi.org/10.1364/JOSAA.29.000313}
{doi: {{%
10\hspace{.1pt}\discretionary{.}{%
}{.}\hspace{.4pt}1364\discretionary{/}{%
}{/}JOSAA\hspace{.1pt}\discretionary{.}{%
}{.}\hspace{.4pt}29\hspace{.1pt}\discretionary{.}{%
}{.}\hspace{.4pt}000313}}}


\bibitem{virtualhikeplan:bleisch:08}
S.~Bleisch and J.~Dykes.
\newblock Using web-based 3-d visualization for planning hikes virtually: An
  evaluation.
\newblock In {\em Representing, Modeling, and Visualizing the Natural
  Environment}, pp. 353--365. 12 2008.
  \href{https://doi.org/10.1201/9781420055504.ch21}
{doi: {{%
10\hspace{.1pt}\discretionary{.}{%
}{.}\hspace{.4pt}1201\discretionary{/}{%
}{/}9781420055504\hspace{.1pt}\discretionary{.}{%
}{.}\hspace{.4pt}ch21}}}


\bibitem{pisteprofile:bortnyk:20}
S.~Bortnyk, N.~Pohorilchuk, O.~Kovtoniuk, and N.~Korohoda.
\newblock {The use of GIS in the morphological analysis of pistes}.
\newblock {\em Geoinformatics: Theor. Appl. Aspects}, (1):1--5, 2020.
  \href{https://doi.org/10.3997/2214-4609.2020geo104}
{doi: {{%
10\hspace{.1pt}\discretionary{.}{%
}{.}\hspace{.4pt}3997\discretionary{/}{%
}{/}2214\discretionary{%
}{-}{-}4609\hspace{.1pt}\discretionary{.}{%
}{.}\hspace{.4pt}2020geo104}}}


\bibitem{slopesapp}
{Breakpoint Studio}.
\newblock {Slopes Ski \& Snowboard - Track Your Winter Adventures}.
\newblock \url{https://getslopes.com/}, 2023.
\newblock Accessed 27.06.2023.

\bibitem{cartography:buckley:04}
A.~Buckley, L.~Hurni, K.~Kriz, T.~Patterson, and J.~Olsenholler.
\newblock Cartography and visualization in mountain geomorphology.
\newblock {\em Geogr. Inf. Sci. Mountain Geomorphol.}, 01 2004.

\bibitem{carden2017simply}
T.~Carden and P.~M. Salmon.
\newblock {Simply Complex: Are LED Outdoor Activities Complex Sociotechnical
  Systems?}
\newblock In P.~Salmon and A.-C. Macquet, eds., {\em Adv. Human Factors Sports
  Outdoor Recreat.}, pp. 141--152. Springer Int. Publishing, 2017.

\bibitem{chaze2008headinjuries}
B.~Chaze and P.~McDonald.
\newblock {Head Injuries in Winter Sports: Downhill Skiing, Snowboarding,
  Sledding, Snowmobiling, Ice Skating and Ice Hockey}.
\newblock {\em Neurologic Clinics}, 26(1):325--332, 02 2008.
  \href{https://doi.org/10.1016/j.ncl.2007.11.009}
{doi: {{%
10\hspace{.1pt}\discretionary{.}{%
}{.}\hspace{.4pt}1016\discretionary{/}{%
}{/}j\hspace{.1pt}\discretionary{.}{%
}{.}\hspace{.4pt}ncl\hspace{.1pt}\discretionary{.}{%
}{.}\hspace{.4pt}2007\hspace{.1pt}\discretionary{.}{%
}{.}\hspace{.4pt}11\hspace{.1pt}\discretionary{.}{%
}{.}\hspace{.4pt}009}}}


\bibitem{opensnow}
{Cloudnine Weather LLC}.
\newblock {OpenSnow: Weather Forecasts | Snow Reports \& Conditions}.
\newblock \url{https://www.opensnow.com/}, 2023.
\newblock Accessed 27.06.2023.

\bibitem{skigroups:delibasic:22}
B.~Delibašić, S.~Radovanović, M.~Z. Jovanović, Z.~Obradović, M.~Suknović,
  and R.~Lojić.
\newblock {A study on ski groups size and their relationship to the risk of
  injury}.
\newblock {\em Proc. Inst. Mech. Eng., Part P: J. Sports Eng. Technol.}, p.
  17543371221118193, 08 2022. \href{https://doi.org/10.1177/17543371221118193}
{doi: {{%
10\hspace{.1pt}\discretionary{.}{%
}{.}\hspace{.4pt}1177\discretionary{/}{%
}{/}17543371221118193}}}


\bibitem{urbanva:Deng:2023}
Z.~Deng, D.~Weng, S.~Liu, Y.~Tian, M.~Xu, and Y.~Wu.
\newblock {A survey of urban visual analytics: Advances and future directions}.
\newblock {\em Comput. Visual Media}, 9(1):3--39, 2023.
  \href{https://doi.org/10.1007/s41095-022-0275-7}
{doi: {{%
10\hspace{.1pt}\discretionary{.}{%
}{.}\hspace{.4pt}1007\discretionary{/}{%
}{/}s41095\discretionary{%
}{-}{-}022\discretionary{%
}{-}{-}0275\discretionary{%
}{-}{-}7}}}


\bibitem{tourplanning:deng:2022}
Z.~Deng, D.~Weng, and Y.~Wu.
\newblock {You are experienced: Interactive tour planning with crowdsourcing
  tour data from web}.
\newblock {\em J. Visualization}, pp. 1--17, 2022.

\bibitem{allaboard:DiLorenzo:16}
G.~Di~Lorenzo, M.~Sbodio, F.~Calabrese, M.~Berlingerio, F.~Pinelli, and
  R.~Nair.
\newblock {AllAboard: Visual Exploration of Cellphone Mobility Data to Optimise
  Public Transport}.
\newblock {\em IEEE Trans. on Vis. and Comp. Graph.}, 22(2):1036--1050, 2016.
  \href{https://doi.org/10.1109/TVCG.2015.2440259}
{doi: {{%
10\hspace{.1pt}\discretionary{.}{%
}{.}\hspace{.4pt}1109\discretionary{/}{%
}{/}TVCG\hspace{.1pt}\discretionary{.}{%
}{.}\hspace{.4pt}2015\hspace{.1pt}\discretionary{.}{%
}{.}\hspace{.4pt}2440259}}}


\bibitem{dijkstra}
E.~W. Dijkstra.
\newblock {A Note on Two Problems in Connexion with Graphs}.
\newblock {\em Numerische Mathematik}, 1(1):269--271, 1959.

\bibitem{pistepreferences}
M.~Dunlop, B.~Elsey, and M.~Masters.
\newblock {Dynamic Visualisation of Ski Data: A Context Aware Mobile Piste
  Map}.
\newblock {\em ACM Int. Conf. Proc. Series}, pp. 375--378, 01 2007.
  \href{https://doi.org/10.1145/1377999.1378040}
{doi: {{%
10\hspace{.1pt}\discretionary{.}{%
}{.}\hspace{.4pt}1145\discretionary{/}{%
}{/}1377999\hspace{.1pt}\discretionary{.}{%
}{.}\hspace{.4pt}1378040}}}


\bibitem{copernicus}
{European Space Agency}.
\newblock {Copernicus DEM - Global and European Digital Elevation Model
  (COP-DEM)}, 2021. \href{https://doi.org/10.5270/ESA-c5d3d65}
{doi: {{%
10\hspace{.1pt}\discretionary{.}{%
}{.}\hspace{.4pt}5270\discretionary{/}{%
}{/}ESA\discretionary{%
}{-}{-}c5d3d65}}}


\bibitem{resortlinking:FALK:17}
M.~Falk.
\newblock Gains from horizontal collaboration among ski areas.
\newblock {\em Tourism Manage.}, 60:92--104, 2017.
  \href{https://doi.org/10.1016/j.tourman.2016.11.008}
{doi: {{%
10\hspace{.1pt}\discretionary{.}{%
}{.}\hspace{.4pt}1016\discretionary{/}{%
}{/}j\hspace{.1pt}\discretionary{.}{%
}{.}\hspace{.4pt}tourman\hspace{.1pt}\discretionary{.}{%
}{.}\hspace{.4pt}2016\hspace{.1pt}\discretionary{.}{%
}{.}\hspace{.4pt}11\hspace{.1pt}\discretionary{.}{%
}{.}\hspace{.4pt}008}}}


\bibitem{profitinskiresorts:falk:20}
M.~Falk and R.~Steiger.
\newblock Size facilitates profitable ski lift operations.
\newblock {\em Tourism Econ.}, 26(7):1197--1211, 2020.
  \href{https://doi.org/10.1177/1354816619868117}
{doi: {{%
10\hspace{.1pt}\discretionary{.}{%
}{.}\hspace{.4pt}1177\discretionary{/}{%
}{/}1354816619868117}}}


\bibitem{fatmap}
{FATMAP}.
\newblock {FATMAP 3D: Map \& Guides for Skiing, Hiking and Biking}.
\newblock \url{https://fatmap.com/adventures/}, 2023.
\newblock Accessed 27.06.2023.

\bibitem{ehyd:Austria:22}
{Federal Ministry Republic of Austria}.
\newblock ehyd.
\newblock \url{https://ehyd.gv.at/}, 2023.
\newblock Accessed 27.06.2023.

\bibitem{skiar:fedosov:16}
A.~Fedosov, E.~Niforatos, I.~Elhart, T.~Schneider, D.~Anisimov, and
  M.~Langheinrich.
\newblock {Design and Evaluation of a Wearable AR System for Sharing
  Personalized Content on Ski Resort Maps}.
\newblock In {\em Proc. Int. Conf. Mobile Ubiquitous Multimedia}, pp. 141--152,
  2016. \href{https://doi.org/10.1145/3012709.3012721}
{doi: {{%
10\hspace{.1pt}\discretionary{.}{%
}{.}\hspace{.4pt}1145\discretionary{/}{%
}{/}3012709\hspace{.1pt}\discretionary{.}{%
}{.}\hspace{.4pt}3012721}}}


\bibitem{osmrouting:felicio:2022}
S.~Felício, J.~Hora, M.~C. Ferreira, D.~Abrantes, P.~D. Costa, C.~Dangelo,
  J.~Silva, and T.~Galvão.
\newblock {Handling OpenStreetMap georeferenced data for route planning}.
\newblock {\em Transp. Res. Procedia}, 62:189--196, 2022.
  \href{https://doi.org/10.1016/j.trpro.2022.02.024}
{doi: {{%
10\hspace{.1pt}\discretionary{.}{%
}{.}\hspace{.4pt}1016\discretionary{/}{%
}{/}j\hspace{.1pt}\discretionary{.}{%
}{.}\hspace{.4pt}trpro\hspace{.1pt}\discretionary{.}{%
}{.}\hspace{.4pt}2022\hspace{.1pt}\discretionary{.}{%
}{.}\hspace{.4pt}02\hspace{.1pt}\discretionary{.}{%
}{.}\hspace{.4pt}024}}}


\bibitem{skimetromap}
K.~Field.
\newblock {Gravity is Your Friend but Every Turn is a Leap of Faith: Design and
  Testing a Schematic Map for Ski Resort Trails}.
\newblock {\em Cartogr. J.}, 47(3):222--237, 2010.
  \href{https://doi.org/10.1179/000870410X12849977317444}
{doi: {{%
10\hspace{.1pt}\discretionary{.}{%
}{.}\hspace{.4pt}1179\discretionary{/}{%
}{/}000870410X12849977317444}}}


\bibitem{vgiquality:fonte:17}
C.~Fonte, V.~Antoniou, L.~Bastin, J.~Estima, J.~Jokar~Arsanjani, J.~Laso~Bayas,
  L.~See, and R.~Vatseva.
\newblock {\em {Mapping and the Citizen Sensor}}, chap. {Assessing VGI Data
  Quality}, pp. 137--163.
\newblock {Ubiquity Press}, 09 2017. \href{https://doi.org/10.5334/bbf.g}
{doi: {{%
10\hspace{.1pt}\discretionary{.}{%
}{.}\hspace{.4pt}5334\discretionary{/}{%
}{/}bbf\hspace{.1pt}\discretionary{.}{%
}{.}\hspace{.4pt}g}}}


\bibitem{osmskirouting}
W.~Friedsam, R.~Hieber, A.~Kharitonov, and T.~Rupp.
\newblock {OSM Ski Resort Routing}.
\newblock In {\em ACM SIGSPATIAL}, pp. 11–--14, 2021.
  \href{https://doi.org/10.1145/3474717.3483628}
{doi: {{%
10\hspace{.1pt}\discretionary{.}{%
}{.}\hspace{.4pt}1145\discretionary{/}{%
}{/}3474717\hspace{.1pt}\discretionary{.}{%
}{.}\hspace{.4pt}3483628}}}


\bibitem{vrmaptransform:ghaemi:22}
Z.~Ghaemi, U.~Engelke, B.~Ens, and B.~Jenny.
\newblock Proxemic maps for immersive visualization.
\newblock {\em Cartogr. Geogr. Inf. Sci.}, 49(3):205--219, 2022.
  \href{https://doi.org/10.1080/15230406.2021.2013946}
{doi: {{%
10\hspace{.1pt}\discretionary{.}{%
}{.}\hspace{.4pt}1080\discretionary{/}{%
}{/}15230406\hspace{.1pt}\discretionary{.}{%
}{.}\hspace{.4pt}2021\hspace{.1pt}\discretionary{.}{%
}{.}\hspace{.4pt}2013946}}}


\bibitem{viscompare:gleicher:18}
M.~Gleicher.
\newblock {Considerations for Visualizing Comparison}.
\newblock {\em IEEE Trans. on Vis. and Comp. Graph.}, 24(1):413--423, 2018.
  \href{https://doi.org/10.1109/TVCG.2017.2744199}
{doi: {{%
10\hspace{.1pt}\discretionary{.}{%
}{.}\hspace{.4pt}1109\discretionary{/}{%
}{/}TVCG\hspace{.1pt}\discretionary{.}{%
}{.}\hspace{.4pt}2017\hspace{.1pt}\discretionary{.}{%
}{.}\hspace{.4pt}2744199}}}


\bibitem{preferencebasedvehiclerouting}
C.~Guo, B.~Yang, J.~Hu, C.~Jensen, and L.~Chen.
\newblock Context-aware, preference-based vehicle routing.
\newblock {\em VLDB J.}, 29:1149--1170, 2020.
  \href{https://doi.org/10.1007/s00778-020-00608-7}
{doi: {{%
10\hspace{.1pt}\discretionary{.}{%
}{.}\hspace{.4pt}1007\discretionary{/}{%
}{/}s00778\discretionary{%
}{-}{-}020\discretionary{%
}{-}{-}00608\discretionary{%
}{-}{-}7}}}


\bibitem{neat:han:12}
B.~Han, L.~Liu, and E.~Omiecinski.
\newblock {NEAT: Road Network Aware Trajectory Clustering}.
\newblock In {\em IEEE 32nd Int. Conf. Distrib. Comp. Syst.}, pp. 142--151,
  2012. \href{https://doi.org/10.1109/ICDCS.2012.31}
{doi: {{%
10\hspace{.1pt}\discretionary{.}{%
}{.}\hspace{.4pt}1109\discretionary{/}{%
}{/}ICDCS\hspace{.1pt}\discretionary{.}{%
}{.}\hspace{.4pt}2012\hspace{.1pt}\discretionary{.}{%
}{.}\hspace{.4pt}31}}}


\bibitem{nasatlx}
S.~G. Hart and L.~E. Staveland.
\newblock {Development of NASA-TLX (Task Load Index): Results of Empirical and
  Theoretical Research}.
\newblock Adv. Psychol., pp. 139--183. 1988.
  \href{https://doi.org/10.1016/S0166-4115(08)62386-9}
{doi: {{%
10\hspace{.1pt}\discretionary{.}{%
}{.}\hspace{.4pt}1016\discretionary{/}{%
}{/}S0166\discretionary{%
}{-}{-}4115\discretionary{%
}{(}{(}08\discretionary{)}{%
}{)}62386\discretionary{%
}{-}{-}9}}}


\bibitem{weatherinresorts:haugom:19}
E.~Haugom and I.~Malasevska.
\newblock The relative importance of ski resort- and weather-related
  characteristics when going alpine skiing.
\newblock {\em Cogent Social Sci.}, 5(1):1681246, 2019.
  \href{https://doi.org/10.1080/23311886.2019.1681246}
{doi: {{%
10\hspace{.1pt}\discretionary{.}{%
}{.}\hspace{.4pt}1080\discretionary{/}{%
}{/}23311886\hspace{.1pt}\discretionary{.}{%
}{.}\hspace{.4pt}2019\hspace{.1pt}\discretionary{.}{%
}{.}\hspace{.4pt}1681246}}}


\bibitem{resortprices:haugom:21}
E.~Haugom, I.~Malasevska, and G.~Lien.
\newblock Optimal pricing of alpine ski passes in the case of crowdedness and
  reduced skiing capacity.
\newblock {\em Empirical Econ.}, 61:469--487, 07 2021.
  \href{https://doi.org/10.1007/s00181-020-01872-w}
{doi: {{%
10\hspace{.1pt}\discretionary{.}{%
}{.}\hspace{.4pt}1007\discretionary{/}{%
}{/}s00181\discretionary{%
}{-}{-}020\discretionary{%
}{-}{-}01872\discretionary{%
}{-}{-}w}}}


\bibitem{roadnetworkfocusregion}
J.-H. Haunert and L.~Sering.
\newblock {Drawing Road Networks with Focus Regions}.
\newblock {\em IEEE Trans. on Vis. and Comp. Graph.}, 17(12):2555--2562, 2011.
  \href{https://doi.org/10.1109/TVCG.2011.191}
{doi: {{%
10\hspace{.1pt}\discretionary{.}{%
}{.}\hspace{.4pt}1109\discretionary{/}{%
}{/}TVCG\hspace{.1pt}\discretionary{.}{%
}{.}\hspace{.4pt}2011\hspace{.1pt}\discretionary{.}{%
}{.}\hspace{.4pt}191}}}


\bibitem{heliski:HENDRIKX:16}
J.~Hendrikx, J.~Johnson, and C.~Shelly.
\newblock {Using GPS tracking to explore terrain preferences of heli-ski
  guides}.
\newblock {\em J. Outdoor Recreat. Tourism}, 13:34--43, 2016.
  \href{https://doi.org/10.1016/j.jort.2015.11.004}
{doi: {{%
10\hspace{.1pt}\discretionary{.}{%
}{.}\hspace{.4pt}1016\discretionary{/}{%
}{/}j\hspace{.1pt}\discretionary{.}{%
}{.}\hspace{.4pt}jort\hspace{.1pt}\discretionary{.}{%
}{.}\hspace{.4pt}2015\hspace{.1pt}\discretionary{.}{%
}{.}\hspace{.4pt}11\hspace{.1pt}\discretionary{.}{%
}{.}\hspace{.4pt}004}}}


\bibitem{multibicyclerouting}
J.~Hrncir, P.~Zilecky, Q.~Song, and M.~Jakob.
\newblock {Practical Multicriteria Urban Bicycle Routing}.
\newblock {\em IEEE Trans. Intell. Transp. Syst.}, 18:1--12, 07 2016.
  \href{https://doi.org/10.1109/TITS.2016.2577047}
{doi: {{%
10\hspace{.1pt}\discretionary{.}{%
}{.}\hspace{.4pt}1109\discretionary{/}{%
}{/}TITS\hspace{.1pt}\discretionary{.}{%
}{.}\hspace{.4pt}2016\hspace{.1pt}\discretionary{.}{%
}{.}\hspace{.4pt}2577047}}}


\bibitem{isogeoquality:iso:13}
{Int. Organisation for Standardisation}.
\newblock {ISO19157:2013 Geographic information - Data quality}, 2013.

\bibitem{ijerph18052506}
S.~B. Jackson, K.~T. Stevenson, L.~R. Larson, M.~N. Peterson, and E.~Seekamp.
\newblock {Outdoor Activity Participation Improves Adolescents’ Mental Health
  and Well-Being during the COVID-19 Pandemic}.
\newblock {\em Int. J. Environ. Res. Public Health}, 18(5), 2021.
  \href{https://doi.org/10.3390/ijerph18052506}
{doi: {{%
10\hspace{.1pt}\discretionary{.}{%
}{.}\hspace{.4pt}3390\discretionary{/}{%
}{/}ijerph18052506}}}


\bibitem{mapmatching:Jensen:2009}
C.~S. Jensen and N.~Tradi{\v{s}}auskas.
\newblock {Map Matching}.
\newblock In {\em {Encycl. Database Syst.}}, pp. 1692--1696. 2009.
  \href{https://doi.org/10.1007/978-0-387-39940-9_215}
{doi: {{%
10\hspace{.1pt}\discretionary{.}{%
}{.}\hspace{.4pt}1007\discretionary{/}{%
}{/}978\discretionary{%
}{-}{-}0\discretionary{%
}{-}{-}387\discretionary{%
}{-}{-}39940\discretionary{%
}{-}{-}9\_215}}}


\bibitem{skidaysprediction}
M.~A. King, A.~S. Abrahams, and C.~T. Ragsdale.
\newblock {Ensemble methods for advanced skier days prediction}.
\newblock {\em Expert Syst. Appl.}, 41(4/1):1176--1188, 2014.
  \href{https://doi.org/10.1016/j.eswa.2013.08.002}
{doi: {{%
10\hspace{.1pt}\discretionary{.}{%
}{.}\hspace{.4pt}1016\discretionary{/}{%
}{/}j\hspace{.1pt}\discretionary{.}{%
}{.}\hspace{.4pt}eswa\hspace{.1pt}\discretionary{.}{%
}{.}\hspace{.4pt}2013\hspace{.1pt}\discretionary{.}{%
}{.}\hspace{.4pt}08\hspace{.1pt}\discretionary{.}{%
}{.}\hspace{.4pt}002}}}


\bibitem{skiresortvisitors}
H.~Konu, T.~Laukkanen, and R.~Komppula.
\newblock {Using Ski Destination Choice Criteria to Segment Finnish Ski Resort
  Customers}.
\newblock {\em Tourism Manage.}, 32(5):1096--1105, 2011.
  \href{https://doi.org/10.1016/j.tourman.2010.09.010}
{doi: {{%
10\hspace{.1pt}\discretionary{.}{%
}{.}\hspace{.4pt}1016\discretionary{/}{%
}{/}j\hspace{.1pt}\discretionary{.}{%
}{.}\hspace{.4pt}tourman\hspace{.1pt}\discretionary{.}{%
}{.}\hspace{.4pt}2010\hspace{.1pt}\discretionary{.}{%
}{.}\hspace{.4pt}09\hspace{.1pt}\discretionary{.}{%
}{.}\hspace{.4pt}010}}}


\bibitem{hikingobstacles:korohoda:21}
N.~Korohoda, O.~Kovtoniuk, T.~Kupach, and N.~Pohorilchuk.
\newblock {Use of GIS for determining mountain local obstacles of routes in low
  mountains}.
\newblock {\em Geoinformatics}, 2021(1):1--6, 2021.
  \href{https://doi.org/10.3997/2214-4609.20215521024}
{doi: {{%
10\hspace{.1pt}\discretionary{.}{%
}{.}\hspace{.4pt}3997\discretionary{/}{%
}{/}2214\discretionary{%
}{-}{-}4609\hspace{.1pt}\discretionary{.}{%
}{.}\hspace{.4pt}20215521024}}}


\bibitem{lera2017analysing}
I.~Lera, T.~P{\'e}rez, C.~Guerrero, V.~M. Egu{\'\i}luz, and C.~Juiz.
\newblock {Analysing Human Mobility Patterns of Hiking Activities through
  Complex Network Theory}.
\newblock {\em PloS one}, 12(5):1--19, 2017.
  \href{https://doi.org/10.1371/journal.pone.0177712}
{doi: {{%
10\hspace{.1pt}\discretionary{.}{%
}{.}\hspace{.4pt}1371\discretionary{/}{%
}{/}journal\hspace{.1pt}\discretionary{.}{%
}{.}\hspace{.4pt}pone\hspace{.1pt}\discretionary{.}{%
}{.}\hspace{.4pt}0177712}}}


\bibitem{skierprefs:Leutschner:71}
W.~Leuschner and R.~Herrington.
\newblock {The Skier: His Characteristics and Preferences}.
\newblock {\em For. Recreat. Symp.}, pp. 135--142, 1971.

\bibitem{shuttlebus:Liu:9331263}
Q.~Liu, Q.~Li, C.~Tang, H.~Lin, X.~Ma, and T.~Chen.
\newblock {A Visual Analytics Approach to Scheduling Customized Shuttle Buses
  via Perceiving Passengers’ Travel Demands}.
\newblock In {\em IEEE Visualization Conf. (VIS)}, pp. 76--80, 2020.
  \href{https://doi.org/10.1109/VIS47514.2020.00022}
{doi: {{%
10\hspace{.1pt}\discretionary{.}{%
}{.}\hspace{.4pt}1109\discretionary{/}{%
}{/}VIS47514\hspace{.1pt}\discretionary{.}{%
}{.}\hspace{.4pt}2020\hspace{.1pt}\discretionary{.}{%
}{.}\hspace{.4pt}00022}}}


\bibitem{complayouts:lyi:21}
S.~LYi, J.~Jo, and J.~Seo.
\newblock {Comparative Layouts Revisited: Design Space, Guidelines, and Future
  Directions}.
\newblock {\em IEEE Trans. on Vis. and Comp. Graph.}, 27(2):1525--1535, 2021.
  \href{https://doi.org/10.1109/TVCG.2020.3030419}
{doi: {{%
10\hspace{.1pt}\discretionary{.}{%
}{.}\hspace{.4pt}1109\discretionary{/}{%
}{/}TVCG\hspace{.1pt}\discretionary{.}{%
}{.}\hspace{.4pt}2020\hspace{.1pt}\discretionary{.}{%
}{.}\hspace{.4pt}3030419}}}


\bibitem{weatherpricing:Malasevska:20}
I.~Malasevska, E.~Haugom, A.~Hinterhuber, G.~Lien, and Ørjan Mydland.
\newblock {Dynamic pricing assuming demand shifting: the alpine skiing
  industry}.
\newblock {\em J. Travel Tourism Mark.}, 37(7):785--803, 2020.
  \href{https://doi.org/10.1080/10548408.2020.1835787}
{doi: {{%
10\hspace{.1pt}\discretionary{.}{%
}{.}\hspace{.4pt}1080\discretionary{/}{%
}{/}10548408\hspace{.1pt}\discretionary{.}{%
}{.}\hspace{.4pt}2020\hspace{.1pt}\discretionary{.}{%
}{.}\hspace{.4pt}1835787}}}


\bibitem{mccurdy2010using}
L.~E. McCurdy, K.~E. Winterbottom, S.~S. Mehta, and J.~R. Roberts.
\newblock {Using Nature and Outdoor Activity to Improve Children's Health}.
\newblock {\em Current Problems in Pediatric and Adolescent Health Care},
  40(5):102--117, 2010. \href{https://doi.org/10.1016/j.cppeds.2010.02.003}
{doi: {{%
10\hspace{.1pt}\discretionary{.}{%
}{.}\hspace{.4pt}1016\discretionary{/}{%
}{/}j\hspace{.1pt}\discretionary{.}{%
}{.}\hspace{.4pt}cppeds\hspace{.1pt}\discretionary{.}{%
}{.}\hspace{.4pt}2010\hspace{.1pt}\discretionary{.}{%
}{.}\hspace{.4pt}02\hspace{.1pt}\discretionary{.}{%
}{.}\hspace{.4pt}003}}}


\bibitem{hikingchoices:molokac:22}
M.~Molokáč, J.~Hlaváčová, D.~Tometzová, and E.~Liptáková.
\newblock {The Preference Analysis for Hikers' Choice of Hiking Trail}.
\newblock {\em Sustainability}, 14(11), 2022.
  \href{https://doi.org/10.3390/su14116795}
{doi: {{%
10\hspace{.1pt}\discretionary{.}{%
}{.}\hspace{.4pt}3390\discretionary{/}{%
}{/}su14116795}}}


\bibitem{ecosus:moreno:18}
J.~Moreno-Gené, L.~Sánchez-Pulido, E.~Cristobal-Fransi, and N.~Daries.
\newblock {The Economic Sustainability of Snow Tourism: The Case of Ski Resorts
  in Austria, France, and Italy}.
\newblock {\em Sustainability}, 10(9), 2018.
  \href{https://doi.org/10.3390/su10093012}
{doi: {{%
10\hspace{.1pt}\discretionary{.}{%
}{.}\hspace{.4pt}3390\discretionary{/}{%
}{/}su10093012}}}


\bibitem{onthesnow}
{Mountain News LLC}.
\newblock {Ski And Snow Reports, Webcams, Skiing Reviews | OnTheSnow.com}.
\newblock \url{https://www.onthesnow.com/}, 2023.
\newblock Accessed 27.06.2023.

\bibitem{osm:osm:17}
{OpenStreetMap contributors}.
\newblock {Planet dump retrieved from https://planet.osm.org }.
\newblock \url{ https://www.openstreetmap.org }, 2017.

\bibitem{owen_renee}
R.~Owen, S.~Priest, and A.~Kotze.
\newblock Applying behaviour analysis to team-building in outdoor learning.
\newblock {\em J. Adventure Educ. and Outdoor Learn.}, pp. 1--14, 2022.
  \href{https://doi.org/10.1080/14729679.2022.2127113}
{doi: {{%
10\hspace{.1pt}\discretionary{.}{%
}{.}\hspace{.4pt}1080\discretionary{/}{%
}{/}14729679\hspace{.1pt}\discretionary{.}{%
}{.}\hspace{.4pt}2022\hspace{.1pt}\discretionary{.}{%
}{.}\hspace{.4pt}2127113}}}


\bibitem{pgrouting}
{pgRouting Community}.
\newblock {pgRouting}.
\newblock \url{https://pgrouting.org/}, 2023.
\newblock Accessed 27.06.2023.

\bibitem{realsnow:picketing:10}
C.~M. Pickering and R.~C. Buckley.
\newblock {Climate Response by the Ski Industry: The Shortcomings of Snowmaking
  for Australian Resorts}.
\newblock {\em Ambio}, 39(5/6):430--438, 2010.
  \href{https://doi.org/10.1007/s13280-010-0039-y}
{doi: {{%
10\hspace{.1pt}\discretionary{.}{%
}{.}\hspace{.4pt}1007\discretionary{/}{%
}{/}s13280\discretionary{%
}{-}{-}010\discretionary{%
}{-}{-}0039\discretionary{%
}{-}{-}y}}}


\bibitem{openskimap}
R.~Porter.
\newblock {OpenSkiMap.org}.
\newblock \url{https://openskimap.org/}, 2023.
\newblock Accessed 27.06.2023.

\bibitem{reddanger:pravo:14}
K.~Pravossoudovitch, F.~Cury, S.~G. Young, and A.~J. Elliot.
\newblock {Is red the colour of danger? Testing an implicit red–danger
  association}.
\newblock {\em Ergonomics}, 57(4):503--510, 2014.
  \href{https://doi.org/10.1080/00140139.2014.889220}
{doi: {{%
10\hspace{.1pt}\discretionary{.}{%
}{.}\hspace{.4pt}1080\discretionary{/}{%
}{/}00140139\hspace{.1pt}\discretionary{.}{%
}{.}\hspace{.4pt}2014\hspace{.1pt}\discretionary{.}{%
}{.}\hspace{.4pt}889220}}}


\bibitem{weatherpredictionsurvey}
X.~Ren, X.~Li, K.~Ren, J.~Song, Z.~Xu, K.~Deng, and X.~Wang.
\newblock {Deep Learning-Based Weather Prediction: {A} Survey}.
\newblock {\em Big Data Res.}, 23:100178, 2021.
  \href{https://doi.org/10.1016/j.bdr.2020.100178}
{doi: {{%
10\hspace{.1pt}\discretionary{.}{%
}{.}\hspace{.4pt}1016\discretionary{/}{%
}{/}j\hspace{.1pt}\discretionary{.}{%
}{.}\hspace{.4pt}bdr\hspace{.1pt}\discretionary{.}{%
}{.}\hspace{.4pt}2020\hspace{.1pt}\discretionary{.}{%
}{.}\hspace{.4pt}100178}}}


\bibitem{graphtheory:riaz:11}
F.~Riaz and K.~M. Ali.
\newblock {Applications of Graph Theory in Computer Science}.
\newblock In {\em Int. Conf. Comput. Intell., Commun. Syst. and Netw.}, pp.
  142--145. {IEEE} Comput. Soc., 2011.
  \href{https://doi.org/10.1109/CICSyN.2011.40}
{doi: {{%
10\hspace{.1pt}\discretionary{.}{%
}{.}\hspace{.4pt}1109\discretionary{/}{%
}{/}CICSyN\hspace{.1pt}\discretionary{.}{%
}{.}\hspace{.4pt}2011\hspace{.1pt}\discretionary{.}{%
}{.}\hspace{.4pt}40}}}


\bibitem{knowledgegeneration:sacha:14}
D.~Sacha, A.~Stoffel, F.~Stoffel, B.~C. Kwon, G.~Ellis, and D.~A. Keim.
\newblock {Knowledge Generation Model for Visual Analytics}.
\newblock {\em IEEE Trans. on Vis. and Comp. Graph.}, 20(12):1604--1613, 2014.
  \href{https://doi.org/10.1109/TVCG.2014.2346481}
{doi: {{%
10\hspace{.1pt}\discretionary{.}{%
}{.}\hspace{.4pt}1109\discretionary{/}{%
}{/}TVCG\hspace{.1pt}\discretionary{.}{%
}{.}\hspace{.4pt}2014\hspace{.1pt}\discretionary{.}{%
}{.}\hspace{.4pt}2346481}}}


\bibitem{scoringfns:schmid:21}
J.~Schmid and J.~Bernard.
\newblock {A Taxonomy of Attribute Scoring Functions}.
\newblock In {\em EuroVis Workshop on Visual Analytics (EuroVA)}. The
  Eurographics Association, 2021.
  \href{https://doi.org/10.2312/eurova.20211095}
{doi: {{%
10\hspace{.1pt}\discretionary{.}{%
}{.}\hspace{.4pt}2312\discretionary{/}{%
}{/}eurova\hspace{.1pt}\discretionary{.}{%
}{.}\hspace{.4pt}20211095}}}


\bibitem{ridersguide:schmid:11}
L.~Schmid, T.~Holleczek, and G.~Tr{\"o}ster.
\newblock {RidersGuide: The First Real-Time Navigation System for Ski Slopes}.
\newblock In {\em Eingebettete Systeme}, pp. 71--80. Springer Berlin
  Heidelberg, 2011. \href{https://doi.org/10.1007/978-3-642-16189-6_8}
{doi: {{%
10\hspace{.1pt}\discretionary{.}{%
}{.}\hspace{.4pt}1007\discretionary{/}{%
}{/}978\discretionary{%
}{-}{-}3\discretionary{%
}{-}{-}642\discretionary{%
}{-}{-}16189\discretionary{%
}{-}{-}6\_8}}}


\bibitem{2d3dtrailheadmap:schobesberger:08}
D.~Schobesberger and T.~Patterson.
\newblock {Evaluating the effectiveness of 2D VS. 3D trailhead maps: A map user
  study conducted at Zion National Park, United States}.
\newblock {\em Bull. Soc. Univ. Cartogr.}, 42(1/2):3--8, 2008.

\bibitem{bikeprefs:scott:21}
D.~M. Scott, W.~Lu, and M.~J. Brown.
\newblock {Route choice of bike share users: Leveraging GPS data to derive
  choice sets}.
\newblock {\em J. Transp. Geogr.}, 90:102903, 2021.
  \href{https://doi.org/10.1016/j.jtrangeo.2020.102903}
{doi: {{%
10\hspace{.1pt}\discretionary{.}{%
}{.}\hspace{.4pt}1016\discretionary{/}{%
}{/}j\hspace{.1pt}\discretionary{.}{%
}{.}\hspace{.4pt}jtrangeo\hspace{.1pt}\discretionary{.}{%
}{.}\hspace{.4pt}2020\hspace{.1pt}\discretionary{.}{%
}{.}\hspace{.4pt}102903}}}


\bibitem{weatherimpact:shih:09}
C.~Shih, S.~Nicholls, and D.~F. Holecek.
\newblock {Impact of Weather on Downhill Ski Lift Ticket Sales}.
\newblock {\em J. Travel Res.}, 47(3):359--372, 2009.
  \href{https://doi.org/10.1177/0047287508321207}
{doi: {{%
10\hspace{.1pt}\discretionary{.}{%
}{.}\hspace{.4pt}1177\discretionary{/}{%
}{/}0047287508321207}}}


\bibitem{arlbergmap}
{Ski Arlberg}.
\newblock {Interactive Map}.
\newblock
  \url{https://www.skiarlberg.at/en/Ski-Arlberg/Live-Info/Interactive-map},
  2022.
\newblock Accessed 27.06.2023.

\bibitem{arlberg}
{Ski Arlberg}.
\newblock {The Cradle of Alpine Skiing}.
\newblock \url{https://www.skiarlberg.at/en}, 2022.
\newblock Accessed 27.06.2023.

\bibitem{paretocycling}
Q.~Song, P.~Zilecky, M.~Jakob, and J.~Hrncir.
\newblock {Exploring Pareto Routes in Multi-Criteria Urban Bicycle Routing}.
\newblock In {\em Int. IEEE Conf. Intell. Transp. Syst. (ITSC)}, pp.
  1781--1787, 2014. \href{https://doi.org/10.1109/ITSC.2014.6957951}
{doi: {{%
10\hspace{.1pt}\discretionary{.}{%
}{.}\hspace{.4pt}1109\discretionary{/}{%
}{/}ITSC\hspace{.1pt}\discretionary{.}{%
}{.}\hspace{.4pt}2014\hspace{.1pt}\discretionary{.}{%
}{.}\hspace{.4pt}6957951}}}


\bibitem{climatechange}
R.~Steiger, E.~Posch, G.~Tappeiner, and J.~Walde.
\newblock The impact of climate change on demand of ski tourism - a simulation
  study based on stated preferences.
\newblock {\em Ecol. Econ.}, 170:106589, 2020.
  \href{https://doi.org/10.1016/j.ecolecon.2019.106589}
{doi: {{%
10\hspace{.1pt}\discretionary{.}{%
}{.}\hspace{.4pt}1016\discretionary{/}{%
}{/}j\hspace{.1pt}\discretionary{.}{%
}{.}\hspace{.4pt}ecolecon\hspace{.1pt}\discretionary{.}{%
}{.}\hspace{.4pt}2019\hspace{.1pt}\discretionary{.}{%
}{.}\hspace{.4pt}106589}}}


\bibitem{sterchi2019avalanche}
R.~Sterchi and P.~Haegeli.
\newblock A method of deriving operation-specific ski run classes for avalanche
  risk management decisions in mechanized skiing.
\newblock {\em Nat. Hazards Earth Syst. Sci.}, 19(1):269--285, 2019.
  \href{https://doi.org/10.5194/nhess-19-269-2019}
{doi: {{%
10\hspace{.1pt}\discretionary{.}{%
}{.}\hspace{.4pt}5194\discretionary{/}{%
}{/}nhess\discretionary{%
}{-}{-}19\discretionary{%
}{-}{-}269\discretionary{%
}{-}{-}2019}}}


\bibitem{STIGSDOTTER2011295}
U.~K. Stigsdotter and P.~Grahn.
\newblock {Stressed individuals' preferences for activities and environmental
  characteristics in green spaces}.
\newblock {\em Urban For. Urban Green.}, 10(4):295--304, 2011.
  \href{https://doi.org/10.1016/j.ufug.2011.07.001}
{doi: {{%
10\hspace{.1pt}\discretionary{.}{%
}{.}\hspace{.4pt}1016\discretionary{/}{%
}{/}j\hspace{.1pt}\discretionary{.}{%
}{.}\hspace{.4pt}ufug\hspace{.1pt}\discretionary{.}{%
}{.}\hspace{.4pt}2011\hspace{.1pt}\discretionary{.}{%
}{.}\hspace{.4pt}07\hspace{.1pt}\discretionary{.}{%
}{.}\hspace{.4pt}001}}}


\bibitem{storandt_2012}
S.~Storandt.
\newblock {Route Planning for Bicycles — Exact Constrained Shortest Paths
  Made Practical via Contraction Hierarchy}.
\newblock {\em Proc. Int. Conf. Autom. Plann. Sched.}, 22(1):234--242, 05 2012.
  \href{https://doi.org/0.1609/icaps.v22i1.13495}
{doi: {{%
0\hspace{.1pt}\discretionary{.}{%
}{.}\hspace{.4pt}1609\discretionary{/}{%
}{/}icaps\hspace{.1pt}\discretionary{.}{%
}{.}\hspace{.4pt}v22i1\hspace{.1pt}\discretionary{.}{%
}{.}\hspace{.4pt}13495}}}


\bibitem{strava}
{Strava, Inc}.
\newblock {Strava | Run and Cycling Tracking on the Social Network for
  Athletes}.
\newblock \url{https://strava.com/}, 2023.
\newblock Accessed 27.06.2023.

\bibitem{vancouvercycle:su:10}
J.~G. Su, M.~Winters, M.~Nunes, and M.~Brauer.
\newblock {Designing a route planner to facilitate and promote cycling in Metro
  Vancouver, Canada}.
\newblock {\em Transp. Res. Part A: Policy Pract.}, 44(7):495--505, 2010.
  \href{https://doi.org/10.1016/j.tra.2010.03.015}
{doi: {{%
10\hspace{.1pt}\discretionary{.}{%
}{.}\hspace{.4pt}1016\discretionary{/}{%
}{/}j\hspace{.1pt}\discretionary{.}{%
}{.}\hspace{.4pt}tra\hspace{.1pt}\discretionary{.}{%
}{.}\hspace{.4pt}2010\hspace{.1pt}\discretionary{.}{%
}{.}\hspace{.4pt}03\hspace{.1pt}\discretionary{.}{%
}{.}\hspace{.4pt}015}}}


\bibitem{outofboundsski}
J.~Sykes, J.~Hendrikx, J.~Johnson, and K.~Birkeland.
\newblock {Combining GPS tracking and survey data to better understand travel
  behavior of out-of-bounds skiers}.
\newblock {\em Appl. Geogr.}, 122, 08 2020.
  \href{https://doi.org/10.1016/j.apgeog.2020.102261}
{doi: {{%
10\hspace{.1pt}\discretionary{.}{%
}{.}\hspace{.4pt}1016\discretionary{/}{%
}{/}j\hspace{.1pt}\discretionary{.}{%
}{.}\hspace{.4pt}apgeog\hspace{.1pt}\discretionary{.}{%
}{.}\hspace{.4pt}2020\hspace{.1pt}\discretionary{.}{%
}{.}\hspace{.4pt}102261}}}


\bibitem{skimaps}
A.~Tait.
\newblock {Mountain Ski Maps of North America: Preliminary Survey and Analysis
  of Style}.
\newblock {\em Cartogr. Perspect.}, (67):5–--18, 2010.
  \href{https://doi.org/10.14714/CP67.110}
{doi: {{%
10\hspace{.1pt}\discretionary{.}{%
}{.}\hspace{.4pt}14714\discretionary{/}{%
}{/}CP67\hspace{.1pt}\discretionary{.}{%
}{.}\hspace{.4pt}110}}}


\bibitem{tirla2014digital}
L.~T{\^\i}rla, E.~Matei, R.~Cuculici, I.~Vijulie, and G.~Manea.
\newblock {Digital Elevation Profile: A Complex Tool for the Spatial Analysis
  of Hiking Trails}.
\newblock {\em J. Environ. Tourism Analyses}, 2(1):48, 2014.

\bibitem{skitourismreport:vanat:21}
L.~Vanat.
\newblock {2021 International Report on Snow \& Mountain Tourism}.
\newblock {\em Overview of the Key Industry Figures for Ski Resorts}, 2021.

\bibitem{stravabias:zander:2016}
Z.~S. Venter, V.~Gundersen, S.~L. Scott, and D.~N. Barton.
\newblock {Bias and precision of crowdsourced recreational activity data from
  Strava}.
\newblock {\em Landscape Urban Plann.}, 232, 04 2023.
  \href{https://doi.org/10.1016/j.landurbplan.2023.104686}
{doi: {{%
10\hspace{.1pt}\discretionary{.}{%
}{.}\hspace{.4pt}1016\discretionary{/}{%
}{/}j\hspace{.1pt}\discretionary{.}{%
}{.}\hspace{.4pt}landurbplan\hspace{.1pt}\discretionary{.}{%
}{.}\hspace{.4pt}2023\hspace{.1pt}\discretionary{.}{%
}{.}\hspace{.4pt}104686}}}


\bibitem{businesscomp:wang:18}
Y.~Wang, H.~Haleem, C.~Shi, Y.~Wu, X.~Zhao, S.~Fu, and H.~Qu.
\newblock {Towards Easy Comparison of Local Businesses Using Online Reviews}.
\newblock {\em Comput. Graph. Forum}, 37(3):63--74, 2018.
  \href{https://doi.org/10.1111/cgf.13401}
{doi: {{%
10\hspace{.1pt}\discretionary{.}{%
}{.}\hspace{.4pt}1111\discretionary{/}{%
}{/}cgf\hspace{.1pt}\discretionary{.}{%
}{.}\hspace{.4pt}13401}}}


\bibitem{betterbus:weng:21}
D.~Weng, C.~Zheng, Z.~Deng, M.~Ma, J.~Bao, Y.~Zheng, M.~Xu, and Y.~Wu.
\newblock {Towards Better Bus Networks: A Visual Analytics Approach}.
\newblock {\em IEEE Trans. on Vis. and Comp. Graph.}, 27(2):817--827, 2021.
  \href{https://doi.org/10.1109/TVCG.2020.3030458}
{doi: {{%
10\hspace{.1pt}\discretionary{.}{%
}{.}\hspace{.4pt}1109\discretionary{/}{%
}{/}TVCG\hspace{.1pt}\discretionary{.}{%
}{.}\hspace{.4pt}2020\hspace{.1pt}\discretionary{.}{%
}{.}\hspace{.4pt}3030458}}}


\bibitem{tripcenteredmetrolayout}
H.~Wu, S.~Takahashi, C.~Lin, and H.~Yen.
\newblock {Travel-Route-Centered Metro Map Layout and Annotation}.
\newblock {\em Comput. Graph. Forum}, 31:925--934, 2012.
  \href{https://doi.org/10.1111/j.1467-8659.2012.03085.x}
{doi: {{%
10\hspace{.1pt}\discretionary{.}{%
}{.}\hspace{.4pt}1111\discretionary{/}{%
}{/}j\hspace{.1pt}\discretionary{.}{%
}{.}\hspace{.4pt}1467\discretionary{%
}{-}{-}8659\hspace{.1pt}\discretionary{.}{%
}{.}\hspace{.4pt}2012\hspace{.1pt}\discretionary{.}{%
}{.}\hspace{.4pt}03085\hspace{.1pt}\discretionary{.}{%
}{.}\hspace{.4pt}x}}}


\bibitem{metromapsurvey:wu:20}
H.-Y. Wu, B.~Niedermann, S.~Takahashi, M.~J. Roberts, and M.~Nöllenburg.
\newblock {A Survey on Transit Map Layout – from Design, Machine, and Human
  Perspectives}.
\newblock {\em Comput. Graph. Forum}, 39(3):619--646, 2020.
  \href{https://doi.org/10.1111/cgf.14030}
{doi: {{%
10\hspace{.1pt}\discretionary{.}{%
}{.}\hspace{.4pt}1111\discretionary{/}{%
}{/}cgf\hspace{.1pt}\discretionary{.}{%
}{.}\hspace{.4pt}14030}}}


\bibitem{triplan:Zhang:22}
X.~Zhang, X.~Pang, X.~Wen, F.~Wang, C.~Li, and M.~Zhu.
\newblock {TriPlan: an interactive visual analytics approach for better tourism
  route planning}.
\newblock {\em J. Visualization}, 26:231--248, 2022.
  \href{https://doi.org/10.1007/s12650-022-00861-8}
{doi: {{%
10\hspace{.1pt}\discretionary{.}{%
}{.}\hspace{.4pt}1007\discretionary{/}{%
}{/}s12650\discretionary{%
}{-}{-}022\discretionary{%
}{-}{-}00861\discretionary{%
}{-}{-}8}}}


\bibitem{multirouteplanning:zhu:20}
S.~Zhu.
\newblock {Multi-objective route planning problem for cycle-tourists}.
\newblock {\em Transp. Lett.}, 14(3):298--306, 2020.
  \href{https://doi.org/10.1080/19427867.2020.1860355}
{doi: {{%
10\hspace{.1pt}\discretionary{.}{%
}{.}\hspace{.4pt}1080\discretionary{/}{%
}{/}19427867\hspace{.1pt}\discretionary{.}{%
}{.}\hspace{.4pt}2020\hspace{.1pt}\discretionary{.}{%
}{.}\hspace{.4pt}1860355}}}


\end{thebibliography}









\end{document}